\documentclass[twocolumn,prl,sort&compress,floatfix,longbibliography]{revtex4-1}
\usepackage{amssymb}
\usepackage{amsmath}
\usepackage{bm}
\usepackage[dvipsnames]{xcolor}
\usepackage[sort&compress]{natbib}

\usepackage{graphicx}

\usepackage{color}
\definecolor{red}{rgb}{1,0,0}
\definecolor{blue}{rgb}{0,0,1}

\renewcommand{\thefigure}{\textbf{\arabic{figure}}}

\usepackage{xr}
\newcommand{\red}[1]{\textcolor{black}{#1}}
\newcommand{\green}[1]{\textcolor{black}{#1}}

\begin{document}
\title{A Polymer Model with Epigenetic Recolouring Reveals a Pathway for the {\it de novo} Establishment and 3D organisation of Chromatin Domains}

\author{D. Michieletto$^1$, E. Orlandini$^2$ and D. Marenduzzo$^1$}
\affiliation{$^1$ SUPA, School of Physics and Astronomy, University of 
	Edinburgh, Peter Guthrie Tait Road, Edinburgh, EH9 3FD, UK\\$^2$ Dipartimento di Fisica e Astronomia and Sezione INFN, Universit\`a di Padova, Via Marzolo 8, Padova, Italy.}

\begin{abstract}
	\textbf{One of the most important problems in development is how epigenetic domains can be first established, and then maintained, within cells. To address this question, we propose a framework which couples 3D chromatin folding dynamics, to a ``recolouring'' process modeling the writing of epigenetic marks. Because many intra-chromatin interactions are mediated by bridging proteins, we consider a ``two-state'' model with self-attractive interactions between two epigenetic marks which are alike (either active or inactive). This model displays a first-order-like transition between a swollen, epigenetically disordered, phase, and a compact, epigenetically coherent, chromatin globule. If the self-attraction strength exceeds a threshold, the chromatin dynamics becomes glassy, and the corresponding interaction network freezes. By modifying the epigenetic read-write process according to more biologically-inspired assumptions, our polymer model with recolouring recapitulates the ultrasensitive response of epigenetic switches to perturbations, and accounts for long-lived multi-domain conformations, strikingly similar to the topologically-associating-domains observed in eukaryotic chromosomes.}
\end{abstract}

\maketitle

\section{Introduction}

The word \emph{``epigenetics''} refers to heritable changes in gene expression that occur without alterations of the underlying DNA sequence~\cite{Alberts2014,Probst2009}. It is by now well established that such changes often arise through biochemical modifications occurring on histone proteins while these are bound to eukaryotic DNA to form nucleosomes, the building blocks of the chromatin fiber~\cite{Alberts2014}. These modifications, or ``epigenetic marks'', are currently thought of as forming a ``histone-code''~\cite{Strahl2000}, which ultimately regulates expression~\cite{Kakutani2001}.

It is clear that this histone-code has to be established \emph{de novo} during cell development and inherited after each cell cycle through major genetic events such as replication, mitosis, or cell division~\cite{Turner2002}. A fundamental question in cell biology and biophysics is, therefore, how certain epigenetic patterns are established, and what mechanism can make them heritable. One striking example of epigenetic imprinting is the ``X chromosome inactivation'', which refers to the silencing of one of the two X chromosomes within the nucleus of mammalian female cells -- this is crucial to avoid over-expression of the genes in the X chromosomes, which would ultimately be fatal for the cell. While the choice of which chromosome should be inactivated is stochastic within embryonic stem cells, it is faithfully inherited in differentiated cells~\cite{Nicodemi2007}. 
The inactivation process is achieved, in practice, through the spreading of repressive histone modifications, which turn the chromosome into a transcriptionally silenced Barr body~\cite{Avner2001,Marks2009,Pinter2012}.
This is an example of an ``epigenetic switch'',  a term which generically refers to the up or down-regulation of specific genes in response to, \emph{e.g.}, seasonal changes~\cite{Wood2014,Bratzel2015,Angel2011a}, dietary restrictions~\cite{Hou2016}, aging~\cite{Kenyon2010} or parental imprinting~\cite{Lim2007}.

Although one of the current paradigms of the field is that the epigenetic landscape and 3D genome folding are intimately related~\cite{Barbieri2012,Brackley2013a,Jost2014B,Cortini2016,Dixon2012,Sexton2012,Boettiger2016,Nora2012,Giorgetti2014}, most of the existing biophysical studies incorporating epigenetic \emph{dynamics} have focused on 1-dimensional (1D) or mean field models~\cite{Dodd2007,Sneppen2008,Micheelsen2010,Dodd2011,Hathaway2012,Sneppen2012,Anink-Groenen2014,Jost2014,Zhang2014,Tian2016}. While these models can successfully explain some aspects of the establishment, spreading, and stability of epigenetic marks, they cannot fully capture the underlying 3-dimensional (3D) dynamic organisation of the chromatin.  
This may, though, be a key aspect to consider: for instance, repressive epigenetic modifications are thought to correlate with chromatin compaction~\cite{Alberts2014,Hathaway2012}, therefore it is clear that there must be a strong feedback between the self-regulated organisation of epigenetic marks and the 3D folding of chromatin. In light of this, here we propose a polymer model of epigenetic switches, which directly couples the 3D dynamics of chromatin folding to the 1D dynamics of epigenetics spreading. 

\begin{figure}[t]
	\centering
	\includegraphics[width=0.48\textwidth]{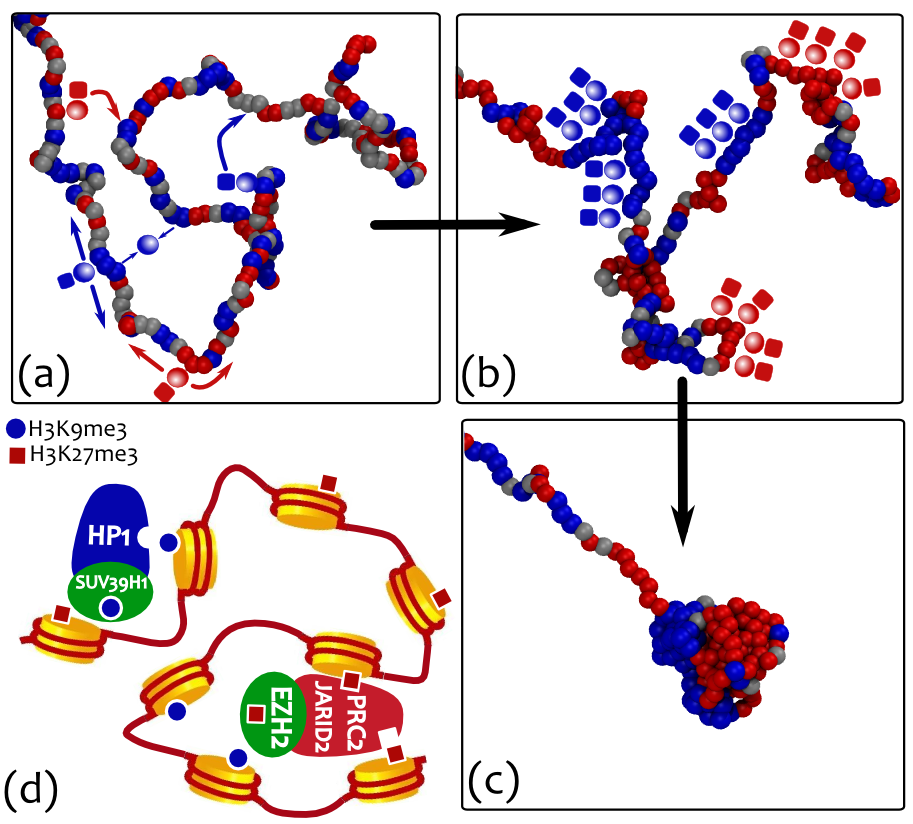}
	\caption{\textbf{A 3D polymer model with ``recolouring'' for the propagation of epigenetic marks.} \textbf{(a)-(c)} Multivalent binding proteins, or ``readers'' (shaded spheres), bind to specific histone modifications and bridge between similarly marked segments (distinguished here via their ``colour''). Histone-modifying enzymes, or ``writers'' (solid squares), are here assumed to be chaperoned by the bridge proteins. The writing (or ``recolouring'') activity is a consequence of 3D contiguity (perhaps through facilitated diffusion~\cite{Brackley2013b}) which is here modeled as a Potts-like interaction between spatially proximate monomers~\cite{Garel1999a} \textbf{(a)}. The positive feedback mechanism and competition between different epigenetic marks results in a regulated spreading of the modifications  \textbf{(b)} which, in turn, drives the overall folding of the polymer \textbf{(c)}. A sketch of a biological reading-writing machinery is shown in \textbf{(d)}. Heterochromatin binding protein HP1 is known to recruit methyltransferase proteins (\emph{e.g.}, SUV39H1) which in turn trimethylates lysine 9 on histone 3 (H3K9me3)~\cite{Peters2001,Hathaway2012,Zentner2013}. Similarly, the Polycomb Repressive Complex (PRC2) is known to comprise histone H3 Lys 27 (H3K27) methyltransferase enzyme EZH2~\cite{Li2010,Angel2011a,Zentner2013} while binding the same mark through the interaction with JARID2~\cite{Li2010,Aranda2015}.}
	\label{fig:model}
\end{figure}

More specifically, we start from the observation that there are enzymes which can either ``read'' or ``write'' epigenetic marks (Fig.~\ref{fig:model}). The ``readers'' are multivalent proteins~\cite{Brackley2013a} which bridge chromatin segments bearing the same histone marks. The ``writers'' are enzymes that are responsible for the establishment and propagation of a specific epigenetic mark, perhaps while performing facilitated diffusion along chromatin~\cite{Brackley2013b}. There is evidence that writers of a given mark are recruited by readers of that same mark~\cite{Dodd2007,Sneppen2008,Lemon2001,Erdel2013,Hathaway2012,Barnhart2011,Angel2011a,Dodd2011}, thereby creating a positive feedback loop which can sustain epigenetic memory~\cite{Sneppen2008}. For example, a region which is actively transcribed by an RNA polymerase is rich in active epigenetic marks (such as the H3K4-methylated marks)~\cite{Lemon2001,Zentner2013}: the polymerase in this example is ``reader'' which recruits the ``writer'' Set1/2~\cite{Ng2003,Zentner2013}. Likewise, the \emph{de novo} formation of centromeres in human nuclei occurs through the creation of the centromere-specific nucleosome CENP-A (a modified histone, which can thus be viewed as an ``epigenetic mark'') via the concerted action of the chaperone protein HJURP (the ``writer'') and the Mis18 complex (the ``reader'')~\cite{Barnhart2011}. Other examples of this read-write mechanism are shown in Fig.~\ref{fig:model}.
This mechanism creates a route through which epigenetic marks can spread to spatially proximate regions on the chromatin, and it is responsible for the coupling between the 3D folding and 1D epigenetic dynamics, addressed for the first time in this work. 

Here we find that, for the simplest case of only 2 epigenetic states which symmetrically compete with each-other (\emph{e.g.}, corresponding to ``active'' or ``inactive'' chromatin~\cite{Alberts2014}), our model predicts a first-order-like phase transition between a swollen, epigenetically disordered, phase, and a collapsed, epigenetically coherent, one. The first-order nature of the transition, within our model, is due to the coupling between 3D and 1D dynamics, and is important because it allows for a bistable epigenetic switch, that can retain memory of its state.
When quenching the system to well below the transition point, we observe a faster 3D collapse of the model chromatin; surprisingly, this is accompanied by a {\it slower} 1D epigenetic dynamics. We call this regime a ``glassy'' phase, which is characterized, in 3D, by a frozen network of strong and short-ranged intra-chain interactions giving rise to dynamical frustration and the observed slowing down, and, in 1D, by a large number of short epigenetic domains.

If the change from one epigenetic mark into the other requires going through an intermediate epigenetic state, we find two main results. First, a long-lived metastable mixed state (MMS), previously absent, is now observed: this is characterized by a swollen configuration of the underlying chain where all epigenetic marks coexist. Second, we find that the MMS is remarkably sensitive to external local perturbations, while the epigenetically coherent states, once established, still display robust stability against major re-organisation events, such as replication. This behaviour is reminiscent of the features associated with epigenetic switches, and the ``X-Chromosome Inactivation'' (XIC). 

We conclude our work by looking at the case in which the epigenetic writing is an ATP-driven, and hence a non-equilibrium process. In this case, detailed balance is explicitly broken and there is no thermodynamic mapping of the underlying stochastic process. This case leads to a further possible regime, characterized by the  formation of a long-lived multi-pearl structure, where each ``pearl'' (or chromatin domain) is associated with a distinct epigenetic domain. This regime is qualitatively different from the glassy phase, as the domains reach a macroscopic size and a significant fraction of chain length. Finally, these self-organised structures are reminiscent of ``topologically associating domains'' (TADs), experimentally observed in chromosomal contact maps~\cite{Lieberman-Aiden2009}. 

\section{Models and Methods}

We model the chromatin fiber as a semi-flexible bead-and-spring chain of $M$ beads of size $\sigma$~\cite{Mirny2011,Rosa2008,Brackley2013a,Barbieri2013,Sanborn2015a,Brackley2016nar}. For concreteness, we consider $\sigma = 3$ kbp $\simeq 30$ nm, corresponding approximately to 15 nucleosomes -- this mapping is commonly used when modeling chromatin dynamics~\cite{Mirny2011,Rosa2008,Brackley2016nar}. 
To each bead, we assign a ``colour'' $q$ representing a possible epigenetic state (mark). Here we consider $q \in \{1,2,3\}$, {\it i.e.} three epigenetic marks such as methylated (inactive), unmarked (intermediate) and acetylated (active). 

\begin{figure*}[t]
	\centering
	\includegraphics[width=1\textwidth]{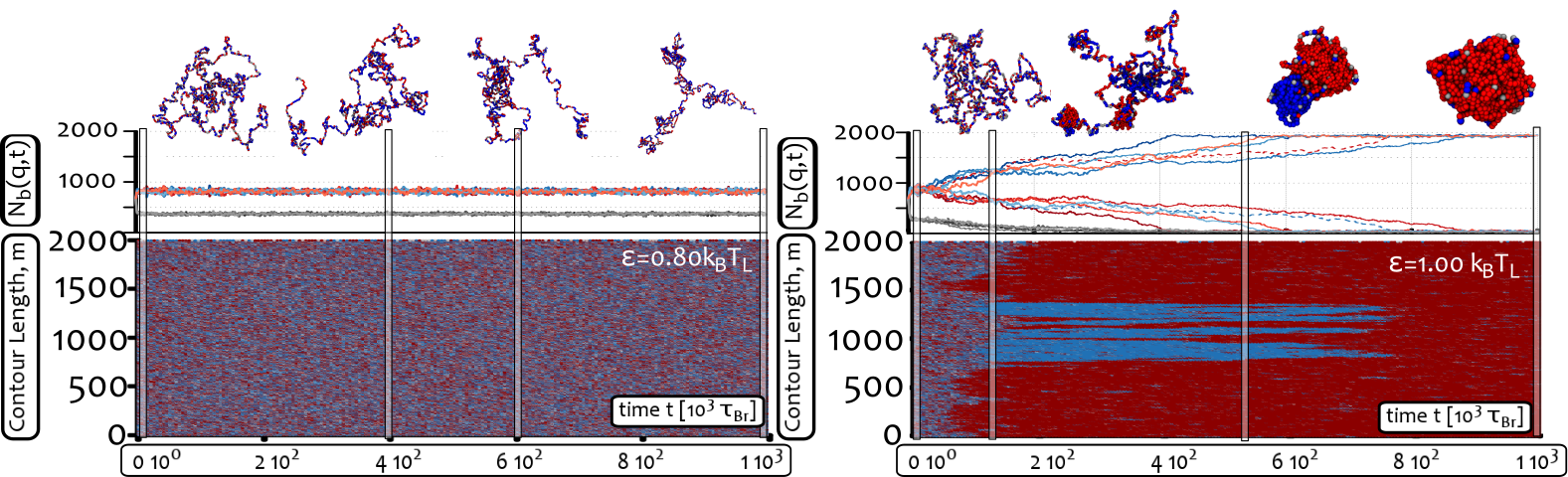}
	\caption{ \textbf{The two-state model above the critical point evolves into an epigenetically coherent state via a symmetry-breaking mechanism.} Top row: typical snapshots of 3D configurations adopted by the polymers as a function of time for two choices of $\alpha = \epsilon/k_BT_L$ below and above the critical point $\alpha_c\simeq 0.9$ (for $M=2000$, see SI). Middle row: time evolution of the total number of beads of type $q$, $N_b(q,t)$, for four independent trajectories (the dashed one corresponds to the trajectory from which the snapshots are taken). Bottom row: time evolution of the colour of each polymer bead, viewed as a ``kymograph''~\cite{Brumley2015}.By tuning $\alpha> \alpha_c$ the whole polymer is taken over by one of the two self-attracting states via a symmetry-breaking mechanism. (see also Suppl. Movies M1-M2).} 
	\label{fig:2Sdb}
\end{figure*}

In addition to the standard effective potentials to ensure chain connectivity (through a harmonic potential between consecutive beads) and bending rigidity (through a Kratky-Porod potential~\cite{Kremer1990}), we consider a repulsive/attractive interaction mediated by the epigenetic marks (colours). This is  described by a truncated-and-shifted Lennard-Jones potential, defined as follows, 
\begin{align}
U_{LJ}^{ab}(x) = \dfrac{4\epsilon_{ab}}{\mathcal{N}} & \left[ \left( \dfrac{\sigma}{x} \right)^{12}  - \left( \dfrac{\sigma}{x} \right)^6  - \left( \dfrac{\sigma}{x_c^{q_aq_b}} \right)^{12} + \left( \dfrac{\sigma}{x_c^{q_aq_b}} \right)^6  \right] \notag \\
&\text{ for } x \leq x^{q_aq_b}_c,
\label{eq:LJ}
\end{align}
whereas $U_{LJ}^{ab}(x) = 0$ for $x > x^{q_aq_b}_c$. In Eq.~(\ref{eq:LJ}), $\mathcal{N}$
is a normalization constant and the parameter $\epsilon_{ab}$ is set so that $\epsilon_{ab}=\epsilon$ for $q_a=q_b$ and $\epsilon_{ab}=k_BT_L$ otherwise. 
The $q$-dependent interaction cut-off $x_c^{q_aq_b}$ is given by $2^{1/6} \sigma$, to model steric repulsion, or $R_i>2^{1/6}\sigma$ to model attraction. [Here, we consider $R_i=1.8\sigma$, which simultaneously ensures short-range interaction and computational efficiency.] In what follows, the cut-offs are chosen so that beads with different colours, or with colour corresponding to no epigenetic marks (i.e., $q=3$), interact via steric repulsion, whereas beads with the same colour, and corresponding to a given epigenetic mark (\emph{e.g.}, $q=1$, or $q=2$), self-attract, modeling interactions mediated by a bridging protein, one of the ``readers''~\cite{Alberts2014,Brackley2013a}.  

The time evolution of the system is obtained by coupling a 3D Brownian polymer dynamics at temperature $T_L$, with a recolouring Monte-Carlo dynamics of the beads which does not conserve the number of monomer types. Recolouring moves are proposed every $\tau_{\rm Rec}=10^3 \tau_{Br}$, where $\tau_{Br}$ is the Brownian time associated with the dynamics of a single polymer bead, and they are realized in practice by attempting $M$ changes of the beads colour. To compare between simulation and physical time units, a Brownian time $\tau_{Br}$ is here mapped to $10$ milliseconds, corresponding to an effective nucleoplasm viscosity $\eta \simeq 150$ cP. This is an intermediate \green{and conservative} value within the range  
that can be estimated from the literature~\cite{Baum2014,Rosa2008} \green{and from a direct mapping with the experimental data of Ref.~\cite{Cabal2006} (see SI Fig.~S1)}. 
With this choice, the recolouring rate is $\sim$ 0.1 s$^{-1}$ and a simulation runtime of $10^6$ Brownian times corresponds to 2.5-3 hours (see SI for more details on the mapping).
Each colour change is accepted according to the standard Metropolis acceptance ratio with effective temperature $T_{\rm Rec}$ and Potts-like energy difference computed between beads that are spatially proximate ({\it i.e.}, within distance $R_i$ in 3D).
It is important to notice that, whenever $T_L \ne T_{\rm Rec}$, detailed balance of the full dynamics is broken, which may be appropriate if epigenetic spreading and writing depend on non-thermal processes (\emph{e.g.}, if they are ATP-driven). More details on the model, and values of all simulation parameters, are given in the SI and Fig.~S1~\footnote{
We should stress at this stage that the recolouring dynamics of epigenetic marks differs from the ``colouring'' dynamics of ``designable'' polymers considered in~\cite{Genzer2012}, where a chemical irreversible patterning is applied for some time to a short polymer in order to study its protein-folded-like conformations~\cite{Genzer2012}. Here, the recolouring dynamics and the folding of the chains evolve together at all times, and they affect one another dynamically. 
}.

The model we use therefore couples an Ising-like (or Potts-like) epigenetic recolouring dynamics, to the 3-dimensional kinetics of polymer folding. In most simulations we consider, for simplicity,  $T_L = T_{\rm Rec}$, and we start from an equilibrated chain configuration in the swollen phase ({\it i.e.}, at very large $T_L$), where beads are randomly coloured with uniform probability. The polymer and epigenetic dynamics is then studied tuning the interaction parameter $\alpha = \epsilon/k_BT_L$ to values near or below the critical  value $\alpha_c$ for which we observe the polymer collapse.

\section{Results}

\subsection{{The ``two-state'' model displays a first-order-like transition which naturally explains both epigenetic memory and bistability}} 

\label{sec:2s}

For simplicity, we focus here on the case in which three states are present, but only two of them ($q=1$, red and $q=2$, blue) are self-attractive, while the third is a neutral state that does not self-attract, but can participate to colouring dynamics ($q=3$, grey). Transition between any two of these three states are possible in this model. Because we find that the grey (unmarked) state rapidly disappears from the polymer at the advantage of the self-attractive ones, we refer to this as an effectively ``two-state'' model. This scenario represents the case with two competing epigenetic marks ({\it e.g.}, an active acetylation mark and an inactive methylation mark), while the third state represents unmarked chromatin. 

Fig.~\ref{fig:2Sdb} reports the polymer and epigenetic dynamics (starting from the swollen and randomly coloured initial state), for two different values of $\alpha = \epsilon/k_BT_L$ below and above the critical point $\alpha_c$. The global epigenetic recolouring is captured by $N_b(q,t)$, the total number of beads in state $q$ at time $t$; the local epigenetic dynamics is instead represented by a ``kymograph''~\cite{Brumley2015}, which describes the change in colour of the polymer beads as time evolves (Fig.~\ref{fig:2Sdb}).

It is readily seen that above the critical point $\alpha_c\simeq 0.9$ (for $M=2000$), the chain condenses fairly quickly into a single globule and clusters of colours emerge and coarsen. Differently-coloured clusters compete, and the system ultimately evolves into an epigenetically coherent globular phase. This is markedly different from the case in which $\alpha<\alpha_c$ where no collapse and epigenetic ordering occurs. Because the red-red and blue-blue interactions are equal, the selection of which epigenetic mark dominates is via symmetry-breaking of the red$\leftrightarrow$blue ($\mathbb{Z}_2$) symmetry. 

\begin{figure}[t]
	\centering
	\includegraphics[width=0.46\textwidth]{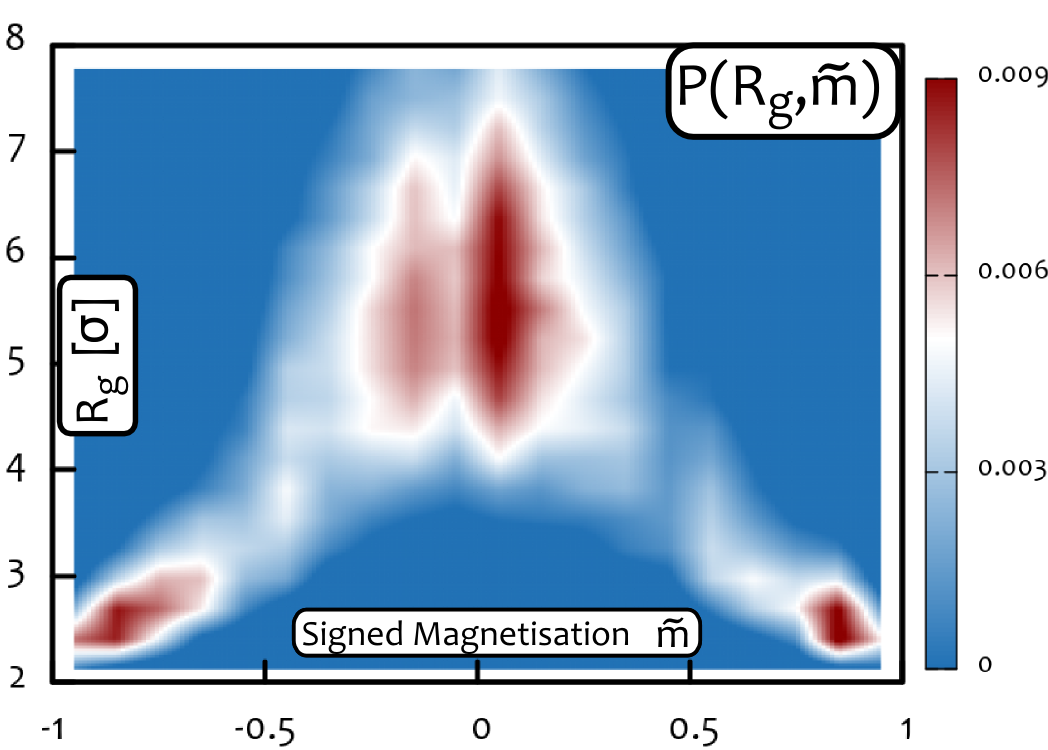}
	\caption{\textbf{The ``two-state'' model displays a discontinuous transition at the critical point marked by coexistence.} Plot of the joint probability $P(R_g,\tilde{m})$ for a chain of $M=50$ beads, obtained from $100$ independent simulations of duration $10^6 \tau_{Br}$ each ($1000$ recolouring steps) at $\alpha=1.15$ (the critical point for $M=50$). Single trajectories are shown in the SI. One can readily appreciate that the system displays coexistence at the critical point, therefore suggesting it is a discontinuous, first-order-like, transition (see SI Fig.~S3 for plots of $P(R_g,\tilde{m})$ at other values of $\alpha$).} 
	\label{fig:2Sdb_criticalpoint}
\end{figure}

The transition between the swollen-disordered and collapsed-coherent phases bears the hallmark of a discontinuous, first-order-like transition~\cite{Grosberg1984,Dormidontova1992}: for instance, we observe metastability of each of the two phases at $\alpha \simeq \alpha_c$ as well as marked hysteresis (see SI, Figs.~S2-S3). To better characterize the transition, we perform a set of simulations on a shorter polymer with $M=50$ beads in order to enhance sampling. We average data from 100 simulations (see SI, Fig.~S4, for single trajectories), each $10^6$ Brownian times long, and calculate the joint probability $P(R_g,\tilde{m})$ of observing a state with a given value of gyration radius, $R_g$, and signed ``epigenetic magnetisation''~\cite{Jost2014}, 
\begin{equation}
\tilde{m} \equiv \dfrac{1}{M} \left( N_b(q=1)-N_b(q=2) \right).
\end{equation}
The result (see Fig.~\ref{fig:2Sdb_criticalpoint} and SI, Fig.~S3) shows that the single maximum expected for the swollen-disordered phase (large $R_g$ and small $\tilde{m}$) splits into two symmetric maxima corresponding to the collapsed-ordered phase (small $R_g$ and $\tilde{m} \simeq \pm 1$). More importantly, at the critical point three maxima are clearly visible suggesting the presence of phase coexistence (see Fig.~\ref{fig:2Sdb_criticalpoint} and SI Fig.~S2-S3). 

The existence of a first-order-like transition in this model provides a marked difference between our model and previous ones, which approximated the epigenetic (recolouring) dynamics as a one-dimensional process, where nucleosome recruitment was regulated by choosing an {\it ad hoc} long-range interaction~\cite{Jost2014,Dodd2007}. These effectively 1D models display either a second order transition~\cite{Dodd2007,Caselle2015,Bouchet2010}, or a first-order transition, but only in the mean-field (``all against all'') case~\cite{Jost2014}. In our model the first-order-nature of the transition critically requires the coupling between the 3D polymer collapse and the 1D epigenetic dynamics -- in this sense, the underlying physics is similar to that of magnetic polymers~\cite{Garel1988}. 

The dynamical feedback between chromatin folding and epigenetic recolouring can be appreciated by looking at Suppl. Movies M1-M2,
where it can be seen that local epigenetic fluctuations trigger local chromatin compaction. Suppl. Movies M1-M2 
also show that the dynamics of the transition from swollen to globular phase is, to some extent, similar to that experienced by a homopolymer in poor solvent conditions~\cite{DeGennes1985b,Kuznetsov1995,Byrne1995,Klushin1998,Kikuchi2002,Kikuchi2005,Ruzicka2012,Leitold2014}. namely a formation of small compact clusters along the chain (pearls) that eventually coalesce into a single globule. 
Unlike the homopolymer case, however, the pearls may be differently coloured giving rise at intermediate or late times to frustrated dynamics, where two or more globules of different colours compete through strong surface tension effects. When several globules are present, we observe cases in which two or more pearls of the same colour, that are distant along the chain but close in 3D, merge by forming long-ranged loops (see snapshots in Fig.~\ref{fig:2Sdb}, contact maps in SI and Suppl. Movies M1-M2). 

Finally, we should like to stress that a first-order-like transition in this system is important for biological applications, since it naturally provides a framework within which epigenetic states can be established and maintained in the presence of external fluctuations. In particular it is well known that when a gene is switched off, for instance after development, it can very rarely be re-activated following further cellular division. This is an example of epigenetic memory, which is naturally explained within our model (as there is hysteresis). At the same time, two cell lines might display different patterns of active and inactive genes, therefore providing a clear example of epigenetic bistability, which is also recovered within this model, due to the red-blue symmetry breaking. All this strongly suggests that the features characterising the above-mentioned ``epigenetic switches'' may be inherited from an effective first-order-like transition driven by the coupling between epigenetic dynamics and chromatin folding as the one displayed by the model presented here.

\begin{figure*}[t]
	\centering
	\includegraphics[width=0.93\textwidth]{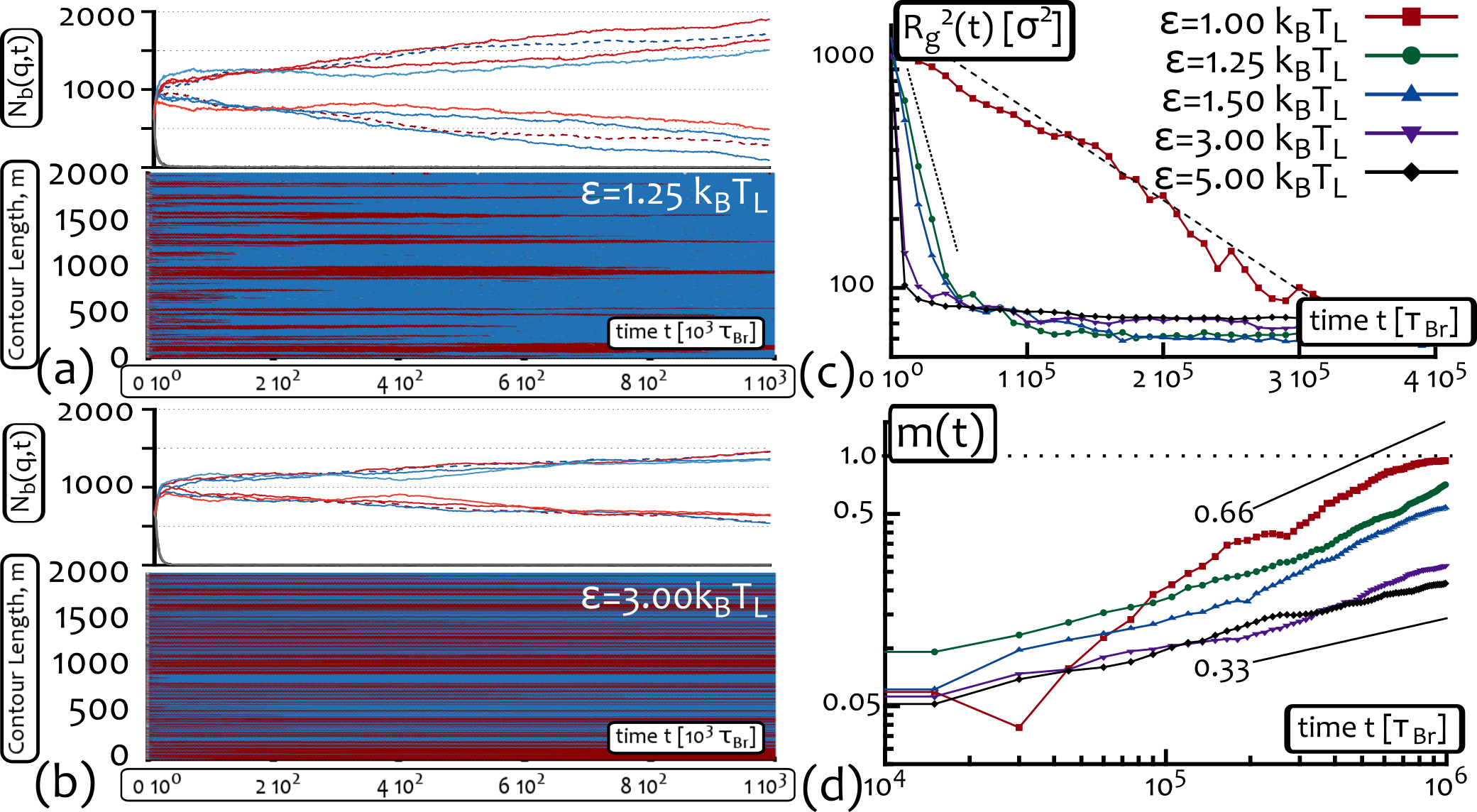}
	\caption{\textbf{Within the two-state model, epigenetic dynamics slows down with increasing $\alpha$}.  \textbf{(a)-(b)} These panels show the kymographs and the number of beads in state $q$, $N_b(q,t)$, for two values of $\alpha$ above the critical point ($\alpha_c\simeq 0.9$ for $M=2000$). Counter-intuitively, the symmetry breaking of the chain towards an epigenetically coherent state slows down with increasing interaction strengths (compare also with Fig.~\ref{fig:2Sdb}). \textbf{(c)} This panel shows the time evolution of the gyration radius $R_g$ of the polymer from the moment the collapse starts. \textbf{(d)} This panel (see also Suppl. Movie M3) shows the behaviour of the epigenetic magnetisation (defined in Eq.~\eqref{eq:magn}) as a function of time. As expected, larger values of $\alpha$ therefore lead to a faster polymer collapse dynamics (faster decay of $R_g$); surprisingly, however, this is accompanied by a slower recolouring dynamics towards the epigenetically coherent state (slower growth of $m(t)$). The longevity of the epigenetic domains thereby formed can be quantified by looking at the growth of the epigenetic magnetisation. For $\alpha=5$, $m(t)$ can be extrapolated to reach, say $0.5$ at about $3$ $10^7$ $\tau_{Br}$ which corresponds to $5000$ minutes of physical time according to our time mapping (see Models and Methods).}
	\label{fig:2Sdb_curves} 
\end{figure*}

\vspace*{-0.0 cm}

\subsection{{Deep quenches into the collapsed phase leads to a ``topological freezing''} \red{which slows down epigenetic dynamics}} 
\label{sec:2sglass}

An intriguing feature observed in the dynamics towards the symmetry-breaking is that quenching at different temperatures affects non trivially the timescales of chromatin condensation and epigenetic evolution towards a single coherent state (see also Suppl. Movie M3). 
The separation between these two timescales increases with $\alpha$ ({\it i.e.}, for deeper quenches), as can be readily seen in Fig~\ref{fig:2Sdb_curves}, where we compare the time evolution of the mean squared radius of gyration of the chain $R_g^2(t)$ and the time-dependent (absolute) epigenetic magnetisation
\begin{equation}
m(t) = \dfrac{1}{M}\left|N_b(q=1,t) - N_b(q=2,t)\right|,
\label{eq:magn}
\end{equation}
for different values of $\alpha$.

While $R_g$ decays exponentially with a timescale that decreases as $\alpha$ increases (Fig.~\ref{fig:2Sdb_curves}(a)),  the epigenetic magnetisation grows as $m(t)\sim t^\beta$, where the dynamical exponent $\beta$ decreases from $\simeq 2/3$ to $\simeq 1/3$ as $\alpha$ increases.
Note that the value $2/3$ has been reported in the literature as the one characterizing the coarsening of pearls in the dynamics of homopolymer collapse~\cite{Byrne1995}. The fact that in our model this exponent is obtained for low values of $\alpha$ suggests that in this regime the timescales of polymer collapse and epigenetic coarsening are similar. In this case, we expect $m(t)$ to scale with the size of the largest pearl in the polymer, whose colour is the most likely to be selected for the final domain -- \emph{i.e.}, the dynamics is essentially determined by the homopolymer case. Our data are instead consistent with an apparent exponent smaller than $2/3$ for larger $\alpha$, signalling a slower epigenetic dynamics.

\begin{figure*}[t]
	\centering
	\includegraphics[width=0.80\textwidth]{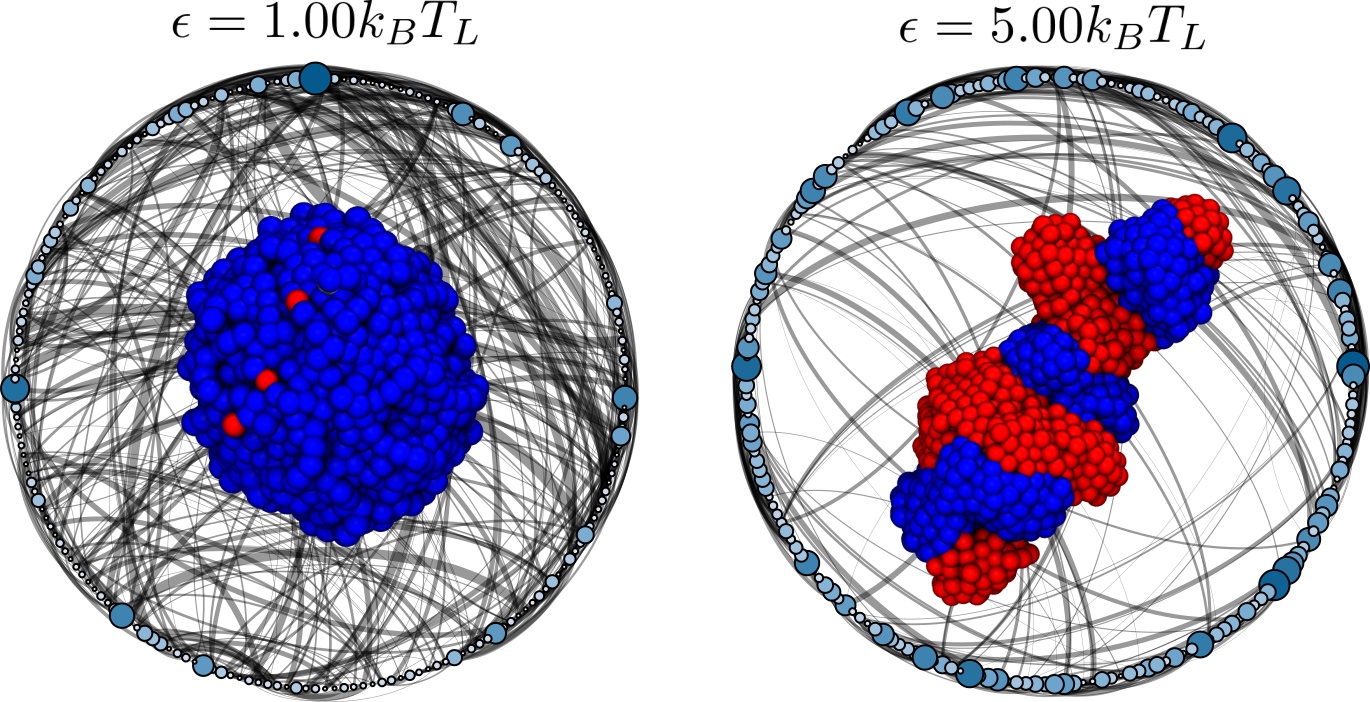}
	\caption{ \textbf{The network of interactions is short ranged for fast collapsing coils.} Snapshot of the network of bead-bead contacts taken at $t=10^6 \tau_{Br}$ for two simulations with (left) $\epsilon=1 k_BT_L$ and (right) $\epsilon=5 k_BT_L$. For clarity of visualization, each node of the network coarse grains 10 beads along the chain. Node size and colour intensity encode the number of interactions within the coarse-grained monomers. Edges are only drawn between nodes which contain interacting monomers, and their thickness is proportional to the (normalized) number of contacts. To improve the visualization, only edges corresponding to a contact probabilities between monomers in the top 30\% are displayed. 
		Snapshots of the respective 3D conformations are also shown. 
		It is important to notice that higher values of $\alpha$ lead to short-ranged networks, which translates in fewer edges but larger nodes in this coarse-grained representation.}
	\label{fig:network_main}
\end{figure*}

The interesting finding that a fast collapse transition gives rise to a slowing down of the recolouring dynamics can be understood in terms of the evolution of the network of intra-chain contacts. This can be monitored by defining the interaction matrix
\begin{equation}
P_{ab}(t)=
\begin{cases} 
1 \text{, if } d_{ab}(t)<R_i \notag\\
0 \text{, otherwise} \notag
\end{cases}
\end{equation} 
where $a,b=1,\ldots,M$ denote two monomers, and $d_{ab}(t)=|{\bm r}_a(t) - {\bm r}_b(t)|$. From the interaction matrix we can readily obtain useful informations on the network structure, such as the average number of neighbours per bead,
\begin{equation}
N_n(t) = \dfrac{1}{M} \sum_{a \neq b} P_{ab}(t)
\end{equation}
or the average ``spanning distance'', which quantifies whether the network is short- or long-ranged (see SI for details). The contact probability between beads $a$ and $b$ can also be simply computed, as the time average of $P_{ab}(t)$.

As expected, for larger values of $\alpha$, $N_n(t)$ saturates to a maximum value (see SI, Fig.~S9).
On the other hand, and more importantly, for higher values of the interaction strength $\alpha$, a dramatic change in the spanning distance is observed. This effect is well captured by plotting a network representation of the monomer-monomer contacts, as reported in Fig.~\ref{fig:network_main} (see SI, Figs.~S6-S9 
for a more quantitative analysis). 
This figure shows that at large $\alpha$ there is a depletion of the number of edges connecting distant monomers along the chain, while short-ranged contacts are enhanced (see caption of Fig.~\ref{fig:network_main} for details; see also contact maps in SI Fig.~S5). 
Note that this finding is consistent with the fractal, or crumpled, globule conjecture~\cite{Mirny2011,Grosberg1988b,Sfatos1997}, for which a globule obtained by a fast collapse dynamics is rich of local contacts and poor in non-local ones. However, the present system represents a novel instance of ``annealed'' collapsing globule, whose segments are dynamically recoloured as it folds.

\begin{figure*}[t]
	\centering
	\includegraphics[width=0.90\textwidth]{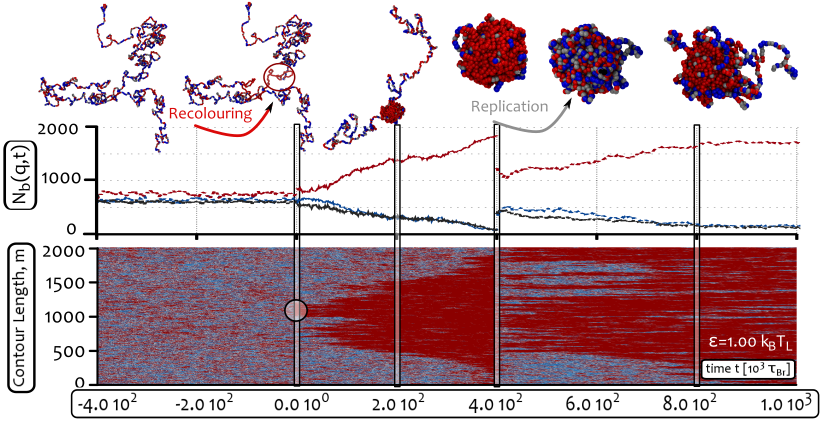}
	\caption{\textbf{The ``two-state with intermediate state'' model displays ultra-sensitive response to external signals such as replication or chromosome inactivation.} Time evolution of the system starting from a mixed metastable state (MMS) and for $\epsilon=k_BT_L$. At $t=0$ a localised perturbation of the MMS is externally imposed by recolouring a segment of 200 beads ($10\%$ of polymer length). This perturbation triggers the collapse of the whole chain into an epigenetically coherent state which is reached within about $4$ $10^5$ Brownian times. At $t=4$ $10^5$ $\tau_{Br}$ we next simulated semi-conservative replication of the collapsed chromatin fiber. This is achieved by assigning a random colour to 50\% of the beads all along the polymer. Following this extensive (\emph{i.e.} non local) colour perturbation, the polymer returns to the epigenetically ordered phase. These results show that the epigenetically coherent phase is robust and stable with respect to extensive perturbations, in stark contrast with the much more sensitive MMS. Suppl. Movie M4 shows the whole dynamics. Contact maps are shown in SI Fig.~S11. 
	}
	\vspace*{- 0.5 cm}
	\label{fig:2S+1db_MMS}
\end{figure*}


Finally, in order to characterize the change in the kinetics of the network, we quantify the ``mobility'' of the contacts, or the ``neighbour exchange rate'', following polymer collapse. We therefore compute
\begin{equation}
\kappa_{n}(t) = \dfrac{1}{M} \sum_{a \neq b} \left[1 - \delta(P_{ab}(t),P_{ab}(t-\Delta t)) \right],  
\end{equation}
where $\Delta t = 10^3 \tau_{Br}=\tau_{\rm Rec}$ is the gap between two measurements. We find that above $\alpha=3$, the time-averaged value of the neighbour exchange rate, normalized by the average number of neighbours, $\langle \kappa_{n} \rangle / \langle N_n \rangle$, sharply drops from values near unity, indicative of mobile rearranging networks, to values close to zero, signalling a frozen network or contacts (see SI Fig.~S10). 

The ``topological freezing'' (see also Suppl. Movie M3)
due to fast folding is also partially reflected by the strongly aspherical shapes taken by the collapsed coils in the large $\alpha$ regime (see snapshots in Fig.~\ref{fig:2Sdb} and Fig.~\ref{fig:network_main}). 

The emerging scenario is therefore markedly different from the one suggested in models for epigenetic dynamics with long-range~\cite{Dodd2007,Caselle2015,Bouchet2010} or mean-field interactions~\cite{Jost2014}, where any two beads in the chain would have a finite interaction probability. Instead, in our case, this is only a valid approximation at small $\alpha$, whereas at large $\alpha$ a given bead interacts with only a subset of other beads (see Fig.~S6), 
and it is only by averaging over different trajectories and beads that we get the power-law decay of the contact probability assumed in those studies~(see Fig.~S7). 
This observation is, once again, intimately related to the fact that we are explicitly taking into account the 3D folding together with the epigenetic dynamics.

In this Section we have therefore shown that considering large interaction strengths between the self-attracting marks (\emph{e.g.} via strongly binding ``readers'') leads to the formation of long-lived and short-ranged domains (see Figs.~\ref{fig:2Sdb_curves}-\ref{fig:network_main} and contact maps in Fig.~S5); while these features might be akin to the ones inferred from experimental contacts maps (Hi-C)~\cite{Lieberman-Aiden2009}, both the network of interactions and the epigenetic dynamics appear to be glassy and frozen (Figs.~\ref{fig:2Sdb_curves} and ~S6-S10) on the timescales of our simulations ($\sim 2.5$-$3$ hours of physical time).

\vspace*{-0.2 cm}

\subsection{{Forcing the passage through the ``unmarked'' state  triggers ultrasensitive kinetic response while retaining a first-order-like transition}}
\label{sec:2s+1}
\vspace*{-0.4 cm}

Up until now, our model has been based on a simple rule for the epigenetic dynamics, where each state can be transformed into any other state. In general, a specific biochemical pathway might be required to change an epigenetic mark~\cite{Alberts2014,Dodd2007}. Often, a nucleosome with a specific epigenetic mark (corresponding to, say, the ``blue'' state), can be  converted into another state (say, the ``red'' one) only after the first mark has been removed.  This two-step re-writing mechanism can be described by considering  a ``neutral'' or ``intermediate'' state (IS) through which any nucleosome has to transit before changing its epigenetic state (say, from ``blue'' to ``red'')~\cite{Dodd2007,Sneppen2012,Micheelsen2010}.
Previous studies, based on mean field or {\it ad hoc} power law interaction rules for the recruitment of epigenetic marks have shown that the presence of such an intermediate unmakred state can enhance bistability and create a long-lived mixed metastable state (MMS), in which all epigenetic states coexist in the same system~\cite{Sneppen2012}.

Differently from the simulations reported in the previous Sections, where we never observed a long-lived mixed state, as the ``red'' or ``blue'' beads rapidly took over the ``grey'' beads, in this case we do observe that the mixed state is metastable for a range of $\alpha \gtrsim \alpha_c$. \green{The observed MMS has a characteristic life-time is much longer than the one observed for the disordered state in the ``two-state'' model when $\alpha \gtrsim \alpha_c$ (see SI, Fig.~S12). }
\green{The observed MMS is reminiscent of the one found in Ref.~\cite{Sneppen2012}, although a difference is the absence of large ordered domains in our case.}


A typical example of a mixed metastable state (MMS) is reported in the early times of Fig.~\ref{fig:2S+1db_MMS}: one can see that it is characterized by a swollen coil with no sign of epigenetic domains, and all three states coexist in the same configuration. To quantify the metastability of the mixed state, we performed 30 independent simulations and found that for $\alpha=1$ the MMS survives with probability $50\%$ after $10^6$ Brownian times. By analysing the survival probability of the MMS as a function of time (see SI, Fig.~S12), we further quantified its characteristic decay time (again at $\alpha=1$) as $1.3$ $10^6$ $\tau_{Br}$, corresponding to about 3 hours in physical time according to our mapping. In contrast, we note that for $\alpha \geq 1.25$ the MMS state is unstable and never observed. 

In order to study the stability of the MMS against external agents, we perturb the system by manually recolouring (in a coherent fashion) a localized fraction (10\%) of beads along the chain. From Fig.~\ref{fig:2S+1db_MMS} one can see that, after the perturbation (performed at $t=0$), the chain forms a nucleation site around the artificially recoloured region that eventually grows as an epigenetically coherent globule. The spreading of the local epigenetic domain throughout the whole chain can be followed from the kymograph in Fig.~\ref{fig:2S+1db_MMS}; it appears that the spreading is approximately linear until the winning mark (here red) takes over the whole chain. The spreading may be linear because the nucleation occurs along an epigenetically disordered swollen chain, so that the mark cannot easily jump long distances along the polymer due to the steep decay for long range contacts in the swollen phase (see also Suppl. Movie M4 and contact maps in Fig.~S11). [Note that the argument for linear spreading also applies to spontaneous nucleation, triggered by a fluctuation rather than by an external perturbation, see SI.] 
The spreading speed can be estimated from the ``wake'' left in the kymograph: it takes $0.4$ $10^6$ Brownian times (about 1 hour of real time) to cover 6 Mbp. 

It is remarkable that, even if the spreading remained linear for a longer polymer, this speed would suffice to spread a mark through a whole chromosome. For instance, the X-chromosome (123 Mbp) could be ``recoloured'' within one cell cycle (24 h). All this suggests that the model presented in this Section may thus be relevant for the fascinating ``X-chromosome inactivation'' in embryonic mammalian cells~\cite{Pinter2012}, and, in more general terms, to the spreading of inactive heterochromatin along chromosomes~\cite{Hathaway2012}.  

It is also worth stressing that, in practice, for an {\it in vivo} chromatin fiber, this local coherent recolouring perturbation might be due to an increase in local concentration of a given ``writer'' (or of a reader-writer pair): our results therefore show that a localised perturbation can trigger an extensive epigenetic response, or ``epigenetic switch'', that might affect a large chromatin region or even an entire chromosome. 

To test the stability of the coherent globular state following the symmetry breaking, we perform an extensive random recolouring of the polymer where one of the three possible states is randomly assigned to 50\% of the beads. This perturbation is chosen because it qualitatively mimics~\green{\footnote{Another strategy that we have tested is to turn 50\% of the beads into inactive, grey, monomers, as this may represent more faithfully what happens immediately after replication, when no histone mark has been deposited yet. The results are nonetheless in qualitative agreement with the ones discussed in the text, since grey beads are non attractive and therefore perturb the system more weakly. We in fact observe that the polymer returns to the collapsed ordered state more quickly in this case with respect to other replication protocols.}}  how epigenetic marks may be semi-conservatively passed on during DNA replication~\cite{Dodd2007,Micheelsen2010,Zerihun2015}. 

After this instantaneous extensive random recolouring (performed at $t=4$ $10^5$ $\tau_{Br}$ in Fig.~\ref{fig:2S+1db_MMS}), we observe that the model chromatin returns to the same ordered state, suggesting that the epigenetically coherent state, once selected, is robust to even extensive perturbations such as semi-conservative replication events (see also Suppl. Movie M4).

The largely asymmetric response of the system against external perturbations, which has been shown to depend on its instantaneous state, is known as ``ultra-sensitivity''~\cite{Sneppen2008}. We have therefore shown that forcing the passage through the ``unmarked'' state triggers ultrasensitivity, while retaining the discontinuous nature of the transition already captured by the simpler ``two state'' model. 

From a physics perspective, the results reported in this Section and encapsulated in Figure~\ref{fig:2S+1db_MMS} are of interest because they show that the presence of the intermediate state do not affect the robustness of the steady states or the nature of the first-order-like transition, therefore the previously discussed main epigenetic features of our model, memory and bistability, are maintained. 

Another important remark is that ultrasensitivity is a highly desirable feature in epigenetic switches and during development. A striking example of this feature is the previously mentioned X-chromosome inactivation in mammalian female embryonic stem cells. While the selection of the chromosome copy to inactivate is stochastic at the embryonic stage, it is important to note that the choice is then epigenetically inherited in committed daughter cells~\cite{Nicodemi2007}. Thus, in terms of the model presented here, one may imagine that a small and localised perturbation in the reading-writing machinery may be able to trigger an epigenetic response that drives a whole chromosome from a mixed metastable state into an inactive heterochromatic state within one cell cycle ({\it e.g.}, an ``all-red'' state in terms on Fig.~\ref{fig:2S+1db_MMS}). When the genetic material is then replicated, an extensive epigenetic fluctuation may be imagined to take place on the whole chromosome. In turn, this extensive (global) perturbation decays over time, therefore leading to the same ``red'' heterochromatic stable state, and ensuring the inheritance of the epigenetic silencing.


\begin{figure*}[t]
	\centering 
	\includegraphics[width=1\textwidth]{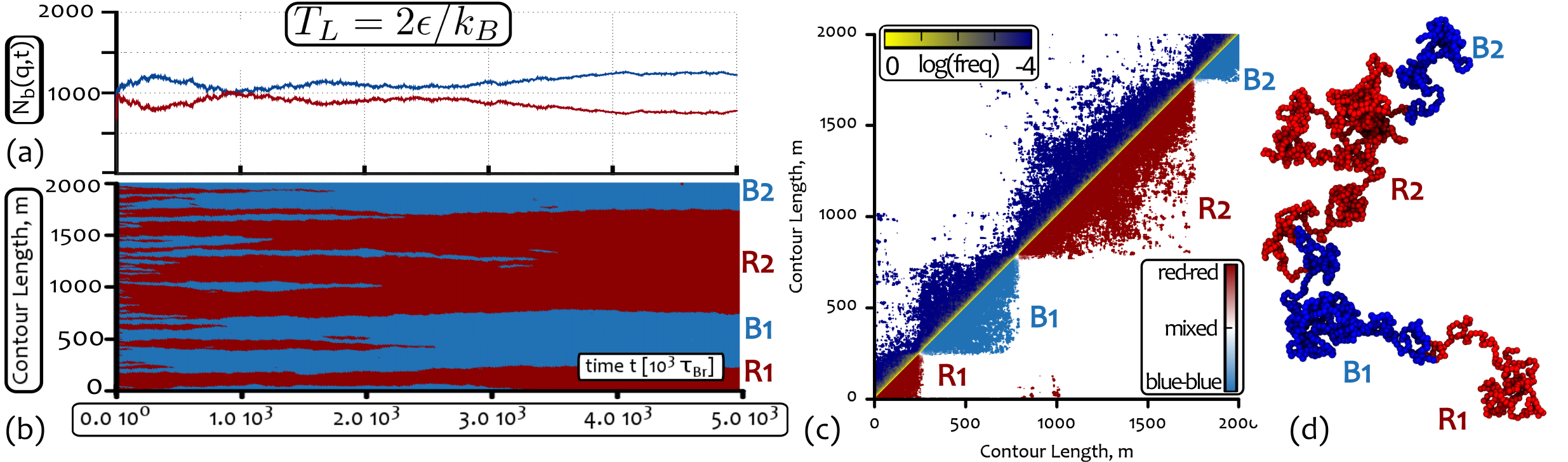}
	\caption{\textbf{Breaking Detailed Balance leads to the formation of TAD-like structures}. Simulations correspond to $M=2000$, $T_{\rm Rec}=0.1 \epsilon/k_B$, $T_L = 2 \epsilon/k_B$ (\emph{i.e.}, $\alpha=\epsilon/k_BT_L=0.5$, see SI for other cases). \textbf{(a)} Plot of the number of red (and blue) coloured beads $N_b(q,t)$ as a function of time. Notice that these curve do not seem to diverge within the simulation runtime, oppositely to the ones reported in the previous Sections. \textbf{(b)} The kymograph of the system showing the presence of long-lived boundaries between distinct epigenetic domains. \textbf{(c)} A contact map averaged over the last $2$ $10^5$ Brownian times: the upper half shows the contact probability between beads, the lower half is colour-coded to separately show the probability of red-red, blue-blue and mixed contacts. \textbf{(d)} A snapshot of the 3D configuration. The visible TAD-like structures in the snapshot and in the contact map are enumerated as in the kymograph, to ease comparison. Note that the TAD-like structures are long-lived but metastable, while coarsening on very long time scales. More details are given in the text and SI, and other values of $T_L$ are given in Figs.~S14-S15 as well as different initial conditions in Fig.~S16. See also Suppl. Movies.}
	\label{fig:2SNOdb}
\end{figure*}

\vspace*{-0.5 cm}

\subsection{Non-equilibrium recolouring dynamics creates a 3D organisation resembling ``topologically associating domains''}
\vspace*{-0.3 cm}

In the previous Sections we have considered the case in which the epigenetic read-write mechanism and the chromatin folding are governed by transition rules between different microstates that obey detailed balance and that can be described in terms of an effective free energy. This is certainly a simplification because the epigenetic writing is in general a non-thermal, out-of-equilibrium process, which entails biochemical enzymatic reactions with chromatin remodelling and ATP consumption~\cite{Alberts2014}. Thus, it is important to see what is the impact of breaking detailed balance in the dynamics of our model.

We address this point by considering a recolouring temperature $T_{\rm Rec}$ that differs from the polymer dynamics temperature $T_L$. When $T_{\rm Rec}\ne T_L$, one can readily show, through the Kolmogorov criterion, that detailed balance is violated, as there is a net probability flux along a closed loop through some of the possible states of the system (see SI). In this case, a systematic scan of the parameter space is computationally highly demanding and outside the scope of the current work.
Here we focus on a specific case where the recolouring temperature is very low, and fixed to $T_{\rm Rec}=0.1 \epsilon/k_BT$, while we vary $T_L$: this case allows to highlight some key qualitative differences in the behaviour of the system which are due to the non-equilibrium epigenetic dynamics. In what follows, we first discuss some expectations based on some general arguments, and then present results from computer simulations.

First, imagine that the Langevin temperature $T_L \to\infty$. In this limit, we expect the polymer to be in the swollen disordered phase, whatever the value of $T_{\rm Rec}$ (no matter how low, as long as greater than zero). This is because a swollen self-avoiding walk is characterized by an intra-chain contact probability scaling as
\begin{equation}\label{scalingPm}
P_c(m) \sim m^{-c}
\end{equation}
with $c=(d+\theta)\nu>2$~\cite{Redner1980,Duplantier1987}. This value implies that the interactions are too short-ranged to trigger a phase transition in the epigenetic state, at least within the Ising-like models considered in Ref.~\cite{Caselle2015}. 

Consider then what happens as $T_L$ decreases. An important lengthscale characterizing order in our system is the epigenetic correlation length, which quantifies the size of the epigenetic domains along the chain. This lengthscale, $\xi$ can be defined through the exponential decay of the epigenetic correlation function (see SI). A second important lengthscale is the blob size. In particular, a homopolymer at temperature $T_L>\Theta$, where $\Theta$ denotes the collapse temperature, can be seen as a collection of transient de Gennes' blobs with typical size~\cite{DeGennes1985b}
\begin{equation}
m^*\sim \left[(T_L-\Theta)/\Theta\right]^{-2}.
\end{equation}
Now, as $T_L$ decreases, remaining larger than $\Theta$, the size of the transient de Gennes' blobs $m^*$ increases. However, these will normally appear randomly along the chain and diffuse over the duration of the simulation to leave no detectable domain in contact maps. If, on the other hand, $\xi \sim m^{*}$, we expect states with one blob per epigenetic domain to be favoured, as the epigenetic recolouring and chromatin folding would be maximally coupled. As a consequence, we may expect the resulting recolouring dynamics to slow down significantly: in this condition, chromatin domains may therefore form, and be long-lived. Finally, the last regime to consider is when $T_L$ is small enough: in this case we expect collapse into an epigenetically coherent globule, similarly to the results from previous Sections.

To test these expectations, we now discuss computer simulations of the ``two-state'' model, where we varied $T_L$ while keeping $T_{\rm Rec}=0.1 \epsilon/k_B$. 
By starting from a swollen disordered polymer (which as previously mentioned is expected to be stable for $T_L\to\infty$), at high enough $T_L$, we find swollen polymers which do not form domains in the simulated contact map (see SI, this phase is also discussed more below). For lower $T_L$ we reach the temperature range that allows for transient blob formation. These are indeed stabilized by the existence of distinct epigenetic domains which appear at the beginning of the simulation; examples of this regime are reported in Fig.~\ref{fig:2SNOdb} and in the SI (Fig.~S15).

This is the most interesting regime as the chromatin fiber displays a multi-pearl structure, reminiscent of the topologically-associating-domains (TADs) found in Hi-C maps~\cite{Lieberman-Aiden2009}. These TADs lead to a ``block-like'' appearance of the contact map (see Figure~\ref{fig:2SNOdb},~\footnote{The coloured contact map is computed by weighting each observed contact with the types of the interacting beads (-1, 1 or 0 for blue-blue, red-red or mixed contacts, respectively) and by normalizing each entry by its total number of contacts. This procedure allows us to identify the epigenetic domains observed in the 3D snapshots, and to demonstrate that in this case the model displays domains which have intra-TAD contacts between same coloured beads, as shown by the roughly uniform colour throughout each domain (more details are given in the SI -- for a full dynamics also see Suppl. Movies M5-M6).}), not unlike the ones reported in the literature~\cite{Dekker2002,Brackley2013a,Brackley2016nar}. Fig.~\ref{fig:2SNOdb} also shows the number of beads in state $q$, $N_b(q,t)$ along with the kymograph tracking the system for $5$ $10^6$ $\tau_{Br}$ timesteps (corresponding to $\sim 14$ hours of physical time according to our mapping). These results show that the boundaries between domains, once established, are long-lived as several are retained throughout the simulation. This figure should be compared and contrasted with Figures~\ref{fig:2Sdb} and \ref{fig:2Sdb_curves}, where the kymographs show either quickly disappearing domains, or long-lived ones that are very small, when the dynamics is glassy. In both those cases, the $N_b(q,t)$ curves show that the system is breaking the red-blue symmetry and the magnetisation is diverging. Here, instead, $N_b(q,t)$ appears to change much more slowly (or is kinetically arrested).

While the TAD-like structure observed at intermediate $T_L$ is long-lived, it might be only metastable, as choosing a swollen but ordered (homopolymer) initial condition, we find that, surprisingly, no domains appear, and the polymer remains homogeneously coloured throughout the simulation without collapsing into a globule. This is a signature of the existence of a swollen but epigenetically ordered phase. We recall that, remarkably, this phase cannot ever be found in the equilibrium limit of the model, $T_L=T_{\rm Rec}$. This new swollen and ordered regime may be due to the fact that, when $T_L$ decreases, the effective contact exponent will no longer be the one for self-avoiding polymers ($c>2$), but it may be effectively closer to the one for ideal ($c=3/2$) or collapsed polymers ($c=1$), both of which allow for long-range interactions between epigenetic segments, possibly triggering epigenetic ordering (see SI, Fig.~S16, ~\footnote{With $T_{\rm Rec}=0.1 \epsilon/k_B$ we cannot simulate large enough $T_L$ to probe the swollen disordered regime: this may signal the fact that $\xi=\infty$ at small enough non-zero $T_{\rm Rec}$, but we cannot exclude it to be a finite size effect (if we simulate polymers with $M<\xi$). On the other hand, the swollen disordered phase can be easily observed at, e.g., $T_{\rm Rec}=0.5 \epsilon/k_B$.}).

Finally, by lowering $T_L$ further, below the theta point for an homopolymer ($T_L \simeq 1.8 \epsilon/k_B$, see SI Fig.~S13) one achieves the point where the polymer collapses into a single epigenetically ordered globule (see SI, Fig.~S15-S16). 

In this Section we have therefore shown that non-equilibrium epigenetic dynamics creates new features in the time evolution and steady state behaviour of the system, and may be important to understand the biophysics of TAD establishment and maintenance. Besides this, we should also mention that the domains emerging in the presented model appear randomly along the chain (\emph{i.e.} no two simulations display the same epigenetic pattern); this is symptomatic of the fact that, for simplicity, our model does not consider structural and insulator elements such as CTCF, promoters, or other architectural~\cite{Alberts2014} and ``bookmarking''~\cite{Sarge2005} proteins which may be crucial for the {\it de novo} establishment of epigenetic domains.
Nonetheless, our model strongly suggests that non-equilibrium processes can play a key role in shaping the organisation of chromosomes. While it has been conjectured for some time that genome regulation entails highly out-of-equilibrium processes, we have here reported a concrete instance in which breaking detailed balance naturally creates a pathway for generating a chromatin organisation resembling the one observed {\it in vivo} chromosomes. 

\section{Discussion and Conclusions}

In this work, we have studied a 3D polymer model with epigenetic ``recolouring'', which explicitly takes into account the coupling between the 3D folding dynamics of a semi-flexible chromatin fiber and the 1D ``epigenetic'' spreading. Supported by several experimental findings and well-established models~\cite{Alberts2014,Brackley2013a}, we assume self-attractive interactions between chromatin segments bearing the same epigenetic mark, but not between unmarked or differently-marked segments. 
We also assume a positive feedback between ``readers'' (binding proteins aiding the folding) and ``writers'' (histone-modifying enzymes performing the recolouring), which is supported by experimental findings and 1D models~\cite{Dodd2007,Sneppen2008,Hathaway2012,Muller-Ott2014,Zentner2013,Aranda2015}.

One important novel element of the presented model is that the underlying epigenetic landscape is \emph{dynamic}, while most of the previous works studying the 3D organisation of chromatin relied on a fixed, or \emph{static}, epigenetic landscape~\cite{Dixon2012,Sexton2012,Boettiger2016,Nora2012,Brackley2013a,Brackley2016nar,Michieletto2016drosophila}. The dynamic nature of the epigenetic modifications is crucial to investigate the {\it de novo} self-organised emergence of epigenetically coherent domains, which is of broad relevance in development and after cell division~\cite{Zentner2013}.  

In particular, the model presented here is able, for the first time to our knowledge, to couple the dynamic underlying epigenetic landscape to the motion of the chromatin in 3D. Furthermore, the synergy between the folding of chromatin and the spreading of histone modifications may be a crucial aspect of nuclear organisation as these two processes are very likely to occur on similar timescales. 
\green{From a biological perspective, one may indeed argue that the formation of local TADs in a cell requires at least several minutes~\cite{Alberts2014}}, while the establishment of higher order, non-local contacts, is even slower~\cite{Michieletto2016drosophila}; at the same time, histone-modifications, such as acetylation or methylation, occur through enzymatic reactions whose rate is of the order of inverse seconds or minutes~\cite{Zentner2013,Barth2010}. For instance, active epigenetic marks are deposited by a travelling polymerase during the $\sim 10$ minutes over which it transcribes an average human gene of 10 kbp~\cite{Cook2001book}.
\green{Similar considerations apply to our work as well: while the microscopic recolouring dynamics takes place over timescales of about $10^3$ $\tau_{\rm Br} \sim 10 s$, the spreading of a coherent mark (e.g. see kymographs in Fig.~\ref{fig:2Sdb},\ref{fig:2Sdb_curves}, \ref{fig:2S+1db_MMS} and \ref{fig:2SNOdb}) may occur on timescales ranging from $5$ $10^5$ $\tau_{\rm Br}$ to $5$ $10^6$ $\tau_{\rm Br}$ which are 5-50 times larger than the polymer re-orientation time (about $10^5$ $\tau_{\rm Br}$, see SI).} 

Furthermore, there are examples of biological phenomena {\it in vivo} which point to the importance of the feedback between 3D chromatin and epigenetic dynamics. A clear example is the inactivation of an active and ``open''~\cite{Alberts2014} chromatin region which is turned into heterochromatin. In this case, the associated methylation marks favour chromatin self-attractive interactions~\cite{Cook2001book} and these, in turn, drive the formation of a condensed structure~\cite{Alberts2014,Zentner2013} whose inner core might be difficult to be reached by other freely diffusing re-activating enzymes.  

Rather fitting in this picture, we highlight that one of our main results is that the coupling between conformational and epigenetic dynamics can naturally drive the transition between a swollen and epigenetically disordered phase at high temperatures and a compact and epigenetically coherent phase at low temperatures (Fig.~\ref{fig:2Sdb}), and that this transition is discontinuous, or first-order-like, in nature (Fig.~\ref{fig:2Sdb_criticalpoint}). 

While it is known that purely short-range interactions 
cannot drive the system into a phase transition, effective (or {\it ad hoc}) 
long-range interactions within an Ising-like framework can induce a (continuous) phase transition in the thermodynamic limit~\cite{Caselle2015,Bouchet2010}. 
In our case, importantly, the transition is discontinuous (see Fig.~\ref{fig:2Sdb_criticalpoint}), and this is intimately related to the coupling between 3D and 1D dynamics. 
The physics leading to a first-order-like transition is therefore reminiscent of that at work for magnetic polymers~\cite{Garel1999a} and hence fundamentally different with respect to previous works, which could not address the conformation-epigenetics positive feedback coupling. 

It is especially interesting to notice that the discontinuous nature of the transition observed in this model can naturally account for bistability and hysteresis, which are both properties normally associated with epigenetic switches. 

We note that the model reported here also displays a richness of physical behaviours. For instance, we intriguingly find that by increasing the strength of self-attraction 
the progress towards the final globular and epigenetically coherent phase is much slower (Fig.~\ref{fig:2Sdb_curves}); we characterize this glass-like dynamics by analysing the network of contacts and identifying a dramatic slowing down in the exchange of neighbours alongside a depletion of non-local contacts (see Figs.~\ref{fig:network_main}). We argue that the physics underlying the emergence of a frozen network of intra-chain interactions might be reminiscent of the physics of spin glasses with quenched disorder~\cite{Garel1997,Sfatos1997,Grosberg1984} (see Figs.~\ref{fig:network_main} and SI Fig.~S10).

We have also shown that the nature of the transition or 
the long-time behaviour of the system is not affected by forcing the passage through an intermediate (neutral or unmarked) state during the epigenetic writing. In contrast, this restriction in kinetic pathway produces major effects on the dynamics. Most notably, it allows for the existence of a long-lived metastable mixed state (MMS) in which all three epigenetic states coexist even above the critical point $\alpha_c$ observed for the simpler ``two-state'' model. 
This case is interesting as it displays ultrasensitivity to external perturbations: the MMS is sensitive to small local fluctuations which drive large conformational and global changes, while the epigenetically coherent states are broadly stable against major and extensive re-organisation events such as semi-conservative chromatin replication (Fig.~\ref{fig:2S+1db_MMS}). 

Like hysteresis and bistability, ultrasensitivity is important in \emph{in vivo} situations, in order to enable regulation of gene expression and ensure heritability of epigenetic marks in development. For instance, it is often that case that, during development, a localized external stimulus (\emph{e.g.}, changes in the concentration of a transcription factor or a morphogen) is enough to trigger commitment of a group of cells to develop into a cell type characterizing a certain tissue rather than another~\cite{Alberts2014}. On the other hand, once differentiated, such cells need to display stability against intrinsic or extrinsic noise. Ultrasensitivity similar to the one we report within this framework would enable both types of responses, depending on the instantaneous chromatin state. 

A further captivating example of ultrasensitive response is the previously mentioned case of the X-chromosome inactivation. Also in that case, the selection of which of the two X-chromosomes to silence is stochastic in female mammalian embryonic stem cells: specifically, it is suggested that a localized increase in the level of some RNA transcripts (XistRNA) can trigger heterochromatization of the whole chromosome, which turns into the so-called Barr body, by propagating repressive marks through recruitment of the polycomb complex PRC2~\cite{Pinter2012}. Once the inactive X copy is selected, the choice is then epigenetically inherited in daughter cells~\cite{Nicodemi2007}, which therefore suggests robustness through disruptive replication events. 

Finally, we have studied the case in which the epigenetic dynamics is subject to a different stochastic noise, with respect to the 3D chromatin dynamics. This effectively ``non-equilibrium'' case, where detailed balance of the underlying dynamics is broken, leads to interesting and unique physical behaviours. Possibly the most pertinent is that we observe, and justify, the existence of a parameter range for which a long-lived multi-pearl state consisting of several globular domains coexist, at least for a time corresponding to our longest simulation timescales which roughly compare to 14 hours of physical time (see Fig.~\ref{fig:2SNOdb} and Models and Methods for the time mapping). This multi-pearl structure is qualitatively reminiscent of the topologically associated domains in which a chromosome folds {\it in vivo}, and requires efficient epigenetic spreading in 1D, together with vicinity to the theta point for homopolymer collapse in 3D. 

Although one of the current paradigms of chromosome biology and biophysics is that the epigenetic landscape directs 3D genome folding~\cite{Barbieri2012,Brackley2013a,Jost2014B,Cortini2016,Boettiger2016}, an outstanding question is how the epigenetic landscape is established in the first place -- and how this can be reset {\it de novo} after each cell division. In this respect, our results suggest that the inherent non-equilibrium (\emph{i.e.}, ATP-driven) nature of the epigenetic read-write mechanism, can provide a pathway to enlarge the possible breadth of epigenetic patterns which can be established stochastically, with respect to thermodynamic models. 

\green{
It is indeed becoming increasingly clear that ATP-driven processes are crucial to regulate chromatin organisation~\cite{Goloborodko2016a,Goloborodko2016}; nonetheless how this is achieved remains largely obscure~\cite{peterbook}. The work presented here provides a concrete example of how this may occur, and suggests that it would be of interest to develop experimental strategies to perturb, for instance, the interaction between reading and writing machines (e.g., by targeting the recruitment between Set1/2 and RNA polymerase, or between EZH2 and PRC, etc.), in order to determine what is the effect of the positive feedback loop on the structure of epigenetic and chromatin domains, and to what extent these require out-of-equilibrium dynamics in order to be established.}

\green{Furthermore, we envisage that the ``recolourable polymer model'' formalised in this work and aimed at studying the interplay between 3D chromatin folding and epigenetic dynamics, might be extended in the future to take into account more biologically detailed (although less general) cases. For instance, one may introduce RNA polymerase as a special ``writer'' of active marks, which can display specific interactions with chromatin, e.g., promote looping~\cite{peterbook}. More generally, our framework can be used as a starting point for a whole family of polymer models which can be used to understand and interpret the outcomes of experiments designed to probe the interplay between \emph{dynamic} epigenetic landscape and chromatin organisation.} 

To conclude, the model presented in this work can therefore be thought of as a general paradigm to study 3D chromatin dynamics coupled to an epigenetic read-write kinetics in chromosomes. All our findings strongly support the hypothesis that positive feedback is a general mechanism through which epigenetic domains, ultrasensitivity and epigenetic switches might be established and regulated in the cell nucleus. We highlight that, within this model, the interplay between polymer conformation and epigenetics plays a major role in the nature and stability of the emerging epigenetic states, which had not previously been appreciated, and we feel ought to be investigated in future experiments. 

We acknowledge ERC for funding (Consolidator Grant THREEDCELLPHYSICS, Ref. 648050). We also wish to thank A.~Y. Grosberg for a stimulating discussion in Trieste. 

\bibliographystyle{apsrev4-1}
\bibliography{PottsOnChain}

\newpage
\setcounter{figure}{0}
\makeatletter 
\renewcommand{\thefigure}{S\@arabic\c@figure}
\makeatother

\section{SUPPLEMENTARY MATERIAL}
\section{Computational Details}
The polymer is simulated as a semi-flexible~\cite{Kremer1990} bead-spring chain 
in which each bead has an internal degree of freedom denoted by $q=\{1,2,3\}$. 

\begin{figure}[t]
	\centering
	\includegraphics[width=0.45\textwidth]{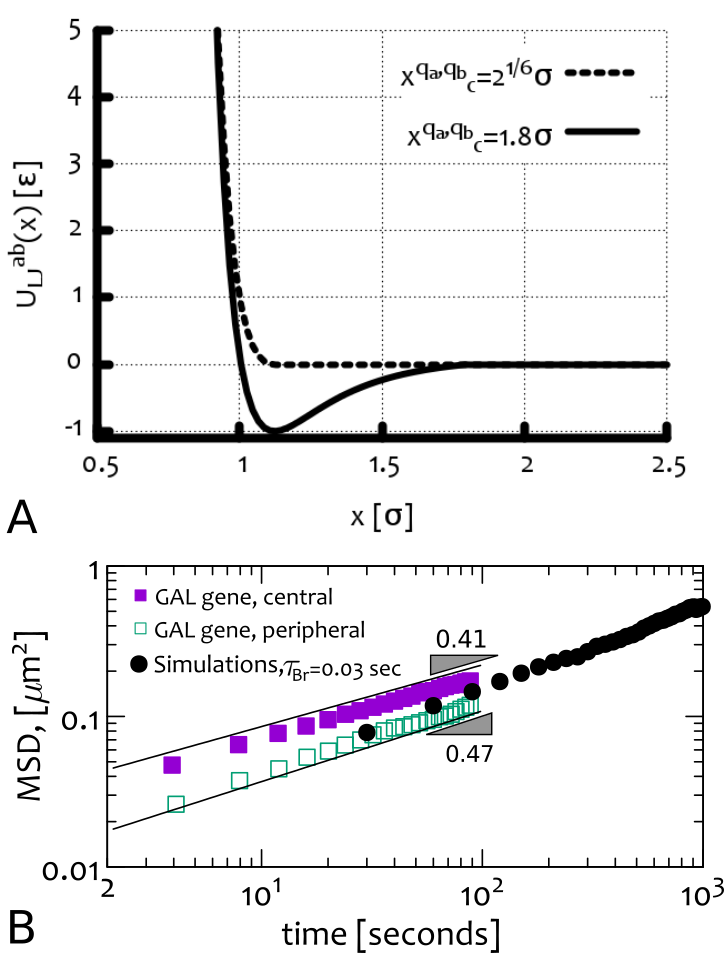}
	\caption{\textbf{Details of the model.}
\textbf{(A)} Shape of the truncated and shifted LJ potential for cut-off $x^{q_a,q_b}_c=1.8\sigma$ (when $q_a=q_b$) and $x^{q_a,q_b}_c=2^{1/6}\sigma$ (when $q_a\neq q_b$).\green{ \textbf{(B)} Direct time mapping of the Brownian time obtained by overlaying simulation data (computed as the mean squared displacement of a polymer bead, averaged over beads and simulations) for $\epsilon=0.9k_BT_L$ and $M=2000$, with experimental data obtained by tracking GAL gene in Yeast~\cite{Cabal2006} (either when it is close to the centre of the nucleus or when localised near the periphery). The best value of $\tau_{\rm Br}$ that matches simulation and experimental data lies around $\tau_{\rm Br}\simeq 0.01-0.05$ seconds.}}
	\label{fig:modLJ}
\end{figure}

The attraction/repulsion between the beads is regulated by the truncated and shifted Lennard-Jones (LJ) potential  as described in the main text:
\begin{align}
U_{LJ}^{ab}(x) = \dfrac{4\epsilon_{ab}}{\mathcal{N}} & \left[ \left( \dfrac{\sigma}{x} \right)^{12}  - \left( \dfrac{\sigma}{x} \right)^6  - \left( \dfrac{\sigma}{x_c^{q_aq_b}} \right)^{12} + \left( \dfrac{\sigma}{x_c^{q_aq_b}} \right)^6  \right] \notag \\
&\text{ for } x \leq x^{q_aq_b}_c 
\label{eq:LJ}
\end{align}
and $U_{LJ}^{ab}(x) = 0$ for $x > x^{q_aq_b}_c$. The $q$-dependent interaction cut-off $x_c^{q_aq_b}$ is set to: (i) $2^{1/6}\sigma$, modelling only steric interaction between beads with different colours, or with colour corresponding to no epigenetic marks (i.e., $q=3$); (ii) $R_1=1.8\sigma$ between beads with the same colour, and corresponding to a given epigenetic mark (e.g., $q=1$, or $q=2$), modelling self-attraction, e.g., mediated by a bridging protein~\cite{Alberts2014}.  
The free parameter $\epsilon_{ab}$ is set so that $\epsilon_{ab}=\epsilon$ for $q_a=q_b=\{1,2\}$ and $\epsilon_{ab}=k_BT_L$ otherwise. Because the potential is shifted to equal zero at the cut-off, we normalise $U_{LJ}^{ab}(x)$ by $\mathcal{N}$ in order to set the minimum of the attractive part to $-\epsilon$ (see also Fig.~\ref{fig:modLJ}). 

The connectivity is taken into account via a harmonic potential between consecutive beads 
\begin{equation}
U_{harm}^{ab}(x) = \dfrac{k_h}{2}(x-x_0)^2 (\delta_{b,a+1}+\delta_{b,a-1})
\end{equation}
where $x_0 = 2^{1/6} \sigma$ and $k_h=200 \epsilon$.
The stiffness is modelled via a Kratky-Porod term~\cite{Kremer1990}
\begin{equation}
U^{ab}_{KP}(x) = \dfrac{k_BT_L l_K}{2 \sigma} \left[ 1 - \dfrac{{\bm t}_a \cdot {\bm t}_b}{|{\bm t}_a| |{\bm t}_b|} \right] (\delta_{b,a+1} + \delta_{b,a-1})
\end{equation}
where ${\bm t_a}$ and ${\bm t}_b$ are the vectors joining monomers $a$,$a+1$ and $b$,$b+1$ respectively. The parameter $l_K/2$ is identified with the persistence length $l_P$ of the chain, here set to $l_P=3 \sigma$.

The total potential $U^a(x)$ experienced by each bead is given by the sum over all the possible interacting pairs and triplets, \emph{i.e.}
\begin{equation}
U^a(x) = \sum_{b\neq a} \left( U^{ab}_{LJ}(x) + U_{harm}^{ab}(x) +  U^{ab}_{KP}(x) \right).
\end{equation} 

The dynamics of each bead is evolved by means of a Brownian Dynamics (BD) 
scheme, \emph{i.e.} with implicit solvent. The corresponding  Langevin equation reads
\begin{equation}
m \dfrac{d^2 {\bm r}_a}{dt^2} = -\gamma \dfrac{d {\bm r}_a}{dt} - {\bm \nabla} U^a(x)+ {\bm \xi}_a
\end{equation}
where $\gamma$ is the friction coefficient and ${\bm \xi}_a$ a stochastic noise which obeys the fluctuation dissipation relationship $\langle \xi_{a,\alpha}(t) \xi_{b,\beta}(t^\prime) \rangle = 2\gamma k_BT_L \delta_{a,b} \delta(t-t^\prime) \delta_{\alpha, \beta}$, where the Latin indexes run over particles while Greek indexes over Cartesian components. 

Using the Einstein relation we set
\begin{equation}
D= \dfrac{k_BT_L}{\gamma} = \dfrac{k_BT_L}{3\pi\eta\sigma},
\end{equation}
where $\eta$ is the solution viscosity. The effective viscosity of the nucleoplasm depends on particle size and timescales: here we consider a bead size of $\sigma=30$ nm, corresponding to 3 kbp~\cite{Rosa2008,Brackley2013a}. A linear extrapolation from the data in Ref.~\cite{Baum2014} would lead to $\eta \sim 5-10$ cP for the early time viscosity for a particle of size $30$ nm -- this is a lower bound as the early time diffusion coefficient larger than the late time value (equivalently, the early time effective viscosity is lower than the late time value)~\cite{Baum2014}. The effective viscosity can also be inferred indirectly from the mapping done in Ref.~\cite{Rosa2008} to fit yeast data; in this case it can be estimated to be in the range $\eta \simeq 100-200$ cP. By using these numbers and $T_L=300 K$ one can define a Brownian time 
\begin{equation}
\tau_{Br} = \sigma^2/D =  \dfrac{3\pi\eta\sigma^3}{k_BT_L } \simeq 0.3-12 \text{ ms}
\end{equation}	
as the time required for a bead to diffuse its own size. \green{We have also performed a direct mapping using the experimental data in yeast of Ref.~\cite{Cabal2006} and the data obtained from our simulations for polymer $M=2000$ beads long and $\epsilon=0.9k_BT_L$. Comparing the mean square displacement of the monomers we found that, in agreement with the previous discussion, the best match between the datasets is attained for $\tau_{Br} \simeq 10-50$ ms (see Fig.~\ref{fig:modLJ}(B)). For definitiveness, and using the worst-case scenario within this mapping strategy, we will assume $\tau_{Br}=10$ ms throughout the rest of the work (as in Ref.~\cite{Rosa2008}). For comparison, it is also useful to mention and to bear in mind that the typical re-orientation time for a polymer with no attractive interactions and $M=2000$ beads long is about $10^5$ $\tau_{\rm Br}$ within our numerical scheme.}
The dynamics is then evolved using a velocity-Verlet integration within the LAMMPS engine in Brownian dynamics mode (NVT ensemble). The simulation runtime typically encompasses $10^6$ $\tau_{Br}$ and is therefore comparable to $2.5-3$ hours of real time. 

The systems are simulated in a box of linear size $L$ and in the dilute regime (assuming each monomer occupies a cylindrical volume $\pi\sigma^3/4$ one can estimate the volume fraction as $\rho=M\pi\sigma^3/4L^3 \simeq 0.1$\%, for a number of monomers $M=2000$). The box is surrounded by a purely repulsive wall in order to avoid self-interactions through periodic boundaries. The initial configuration is typically that of an ideal random walk in which each bead assumes a random value (colour) $q$. We then run $10^4$ $\tau_{Br}$ timesteps in which the only force field is an increasingly stronger steric soft repulsion between every pair of beads, while their colour is left unaltered. The explicit form of the soft potential we use is 
\begin{equation}
	U^{ij}_{soft}(d)= A\left[ 1+\cos{\dfrac{\pi d_{ij}}{d_c}} \right]
	\label{eq:utot}
\end{equation}
where $d_c=2^{1/6} \sigma$ is the cutoff distance and $A$ the maximum of the potential at $d_{ij}=0$.This ``warm-up'' equilibration run transforms the ideal random walk conformations into one obeying self-avoiding statistics as it removes the overlaps between monomers.

Following this equilibration, we start the main run, typically consisting of $10^6$ $\tau_{Br}$ timesteps, in which $M$ recolouring moves are attempted every $10^3$ $\tau_{Br}$ timesteps. Each recolouring move is accepted or rejected using a Metropolis algorithm, \emph{i.e.} the acceptance probability is given by
\begin{equation}
p(q \rightarrow q^\prime) = {\rm min}\left( 1, e^{-\Delta E/k_BT_{\rm Rec}} \right), \label{eq:metrop}
\end{equation}
where $\Delta E$ is the difference between the new energy (after recolouring) and the old one (before recolouring). \red{The energy appearing in Eq.~\eqref{eq:metrop} is computed from Eq.~\eqref{eq:utot}. In particular, upon recolouring any one bead, the only part of the energy function that changes is the LJ potential (Eq.~\eqref{eq:LJ} and Fig.~\ref{fig:modLJ}), as same coloured beads interact through an attractive potential while differently coloured ones only through the repulsive part of the potential. It is important to note that the temperature appearing in the exponent is the ``recolouring'' temperature $T_{\rm Rec}$, which is not necessarily identical to $T_L$, the temperature used in the Langevin equation for the stochastic noise.}

\red{The total polymer length is taken $M=2000 \sigma \simeq 6$ $10^4$ nm or $6$ Mbp at the $3$ kbp per bead resolution which we use. When probing the nature of the phase transition of the ``two state'' model we decrease the length to $M=50$ and perform 100 independent simulations of $10^6$ $\tau_{Br}$ in order to enhance sampling (as these short chain equilibrate quickly).}

\section{The Detailed Balance is broken when $T_P \neq T_L$.}
According to the Kolmogorov criterion, in a stochastic dynamics satisfying detailed balance the product of the transition rates over any closed loop over some states of the system must not depend on the sense along which we go through the loop~\cite{Ramakrishan2015}. 
This is not in general the case when $T_{\rm Rec} \neq T_{Langevin}$.
To see why this is so, let us imagine a simple case where two loose beads initially of the same colour interact only with the LJ potential, without any chain in between. Imagine further than the beads are initially close to each other and are then moved apart by a thermal fluctuation. This happens with probability $p^{q}_{\rm near \rightarrow far} = \exp{(-\epsilon/k_BT_L)}$. At this stage, a change in the colour of the bead ($q$) occurs with probability $1$, as there is no energy penalty. When the beads have different colours, they can come close to each other still  with probability $1$, as there is now no attraction or penalty in being close together (as long their distance is greater than $2^{1/6}\sigma$). Once they are back together, also  the recolouring move that causes the two beads to have the same $q$ occurs with probability $1$ as this move is energetically favourable. 
Therefore we obtain
\begin{equation}
p_{loop} = \exp{\left(-\dfrac{\epsilon}{k_BT_L}\right)}.
\end{equation}
By performing the loop in the reverse direction ({\it i.e.} change $q$ first, then separate the beads, change back $q$, and finally put the beads back in contact) one instead obtains 
\begin{equation}
p_{loop^{-1}} = \exp{\left(-\dfrac{\epsilon}{k_BT_{\rm Rec}}\right)} \ne p_{loop}.
\end{equation}
The two transition probabilities are equal only if $T_{\rm Rec}=T_L$. In particular, if $T_L > T_{\rm Rec}$ the ``direct'' loop is more likely to happen than its reverse, while the opposite is true if $T_L < T_{\rm Rec}$: detailed balance is therefore violated when $T_L\ne T_{\rm Rec}$. 

\section{Second Virial Coefficient}
Given our interparticle potential, it is straightforward to extract the second virial coefficient $u_2$ by using the Mayer relation and Eq.~\eqref{eq:LJ}~\cite{LeTreut2016}:
\begin{equation}
	u_2^{ab} = - \int d^3 x \left( e^{-\beta U_{LJ}^{ab}(x)}  - 1\right).
\end{equation}
We find that $u^{ab}_2$ is positive ($u_2^{rep}$) for $q_a \neq q_b$ and negative ($u_2^{att}$) when $q_a=q_b$. In particular, we find that $u_2^{rep}\simeq 4.396$ while $u_2^{att}$ ranges from $-9.3$ (for $\epsilon=1 k_BT_L$) to $-400$ (for $\epsilon=5 k_BT_L$).

\begin{figure*}[t]
	\includegraphics[width=0.9\textwidth]{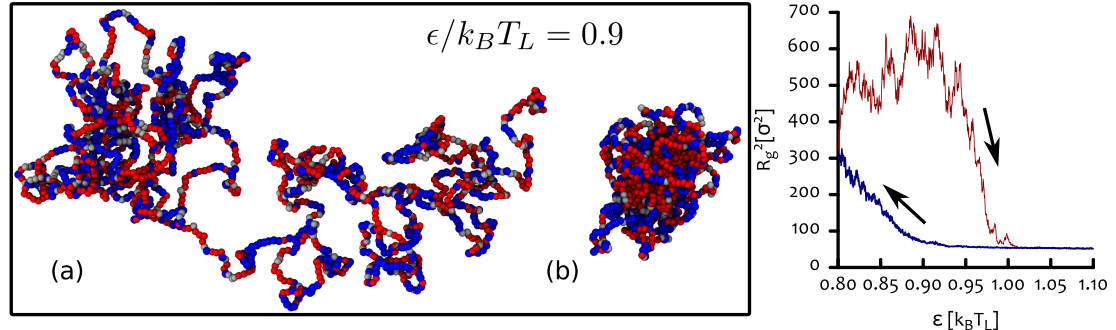}
	\caption{\textbf{Metastability and hysteresis in the two-state model.} 
(a-b) Snapshots corresponding to a chain of $M=2000$ beads in the swollen (a) and globular (b) phase, which are both metastable at the indicated temperature of $\epsilon=0.9 k_BT_L$ -- a simulation starting in one of these phases remain there during a whole run of $10^6$ Brownian times. (c) Plot of the radius of gyration as a function of the interaction strength $\epsilon$ which we slowly increase from $\epsilon=0.8 k_B T_L$ (below the transition) to $\epsilon=1.1 k_B T_L$ (above the transition) in $10^6$ Brownian times (red curve). From there, we decrease the interaction strength back to $\epsilon=0.8 k_B T_L$ in the same amount of time (blue curve). We find that there is a hysteresis cycle, which supports our conclusion that the transition is first-order-like. The curves in (c) are averages over 5 different runs. }
	\label{fig:coexistence}
\end{figure*}

\begin{figure*}[t]
	\includegraphics[width=\textwidth]{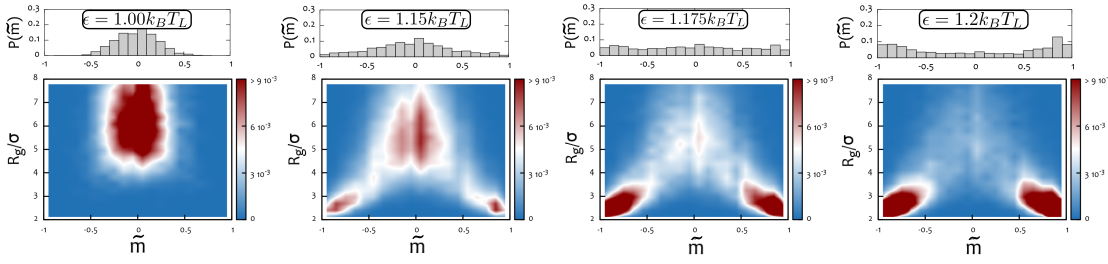}
	\caption{\textbf{First-order-like transition for the two-state model for a polymer with $M=50$.} \green{(Bottom row, from left to right) Heat map representation of the joint probability distribution $P(\tilde{m},R_g)$ of a chain with $M = 50$ and having a radius gyration $R_g$ and a signed epigenetic magnetisation $\tilde{m}$. The four panels refer to the four indicated values of the interaction parameter $\alpha=\epsilon/k_BT_L$ near the critical point. (Top row, from left ro right) By integrating $P(\tilde{m},R_g)$ over $R_g$ one obtains the corresponding reduced distribution $P(\tilde{m})$. As one can see the change from a mono-stable to a bi-stable state below and above the transition point is separated by a state where the distribution is roughly flat.}  Each of the plots is created by averaging over the dynamics of $100$ independent simulations each of duration $10^6$ $\tau_{Br}$ ($1000$ recolouring steps). We stress here that due to finite size effects longer chains display lower values of the critical point $\alpha_c \simeq 0.90$, although we did not thoroughly explore the phase space for the $M=2000$ case (see previous figure).}
	\label{fig:firstorder}
\end{figure*}

\red{
\section{First-Order-Like Nature of the Transition}
We have investigated the nature of the transition from swollen-disordered phase to the collapsed-ordered phase in two ways: (i) by studying hysteresis cycles of a chain with $M=2000$ beads ($5$ runs) and (ii) by measuring the joint probability $P(R_g,\tilde{m})$ from simulations with a well-equilibrated chain with $M=50$ beads ($100$ runs). }

\red{
The results obtained from the first study, (i), are shown in Fig.~\ref{fig:coexistence} (see also Suppl. Movie M7). This figure shows that there is a region of the interaction parameter $\alpha \simeq 0.9-1.0$ for which the two phases (collapsed and swollen) are both metastable. Specifically, $\alpha \simeq 1$ is needed to collapse a swollen chain (red curve), but a lower interaction parameter $\alpha$ is required to send the chain back into the swollen phase, once it is collapsed (blue curve). The curves are made by slowly increasing and decreasing $\epsilon$ over a range of $0.3 k_BT$ over $10^6$ Brownian times.}

\begin{figure*}[t]
	\centering 
	\includegraphics[width=0.9\textwidth]{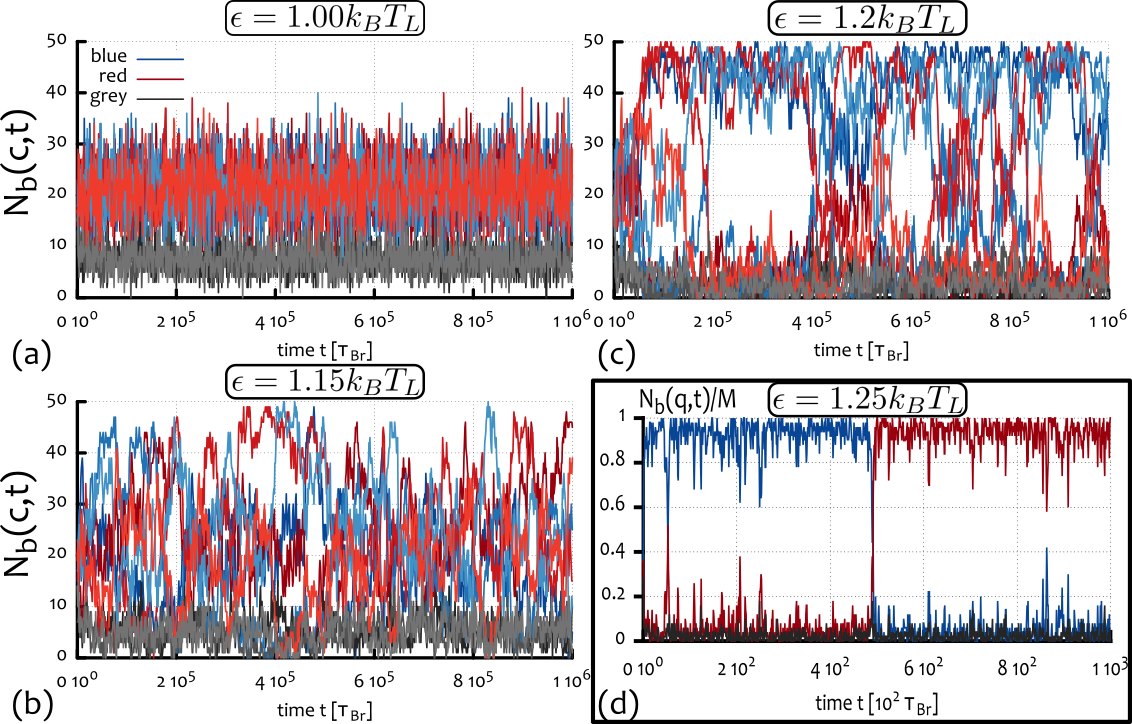}
	\caption{\textbf{Trajectories from simulations with $M=50$ and switch-like behaviour}. \textbf{(a)-(c)} Typical trajectories from simulations of a polymer with $M=50$ beads for three values of $\alpha$ near the critical point. \textbf{(d)} In agreement with previous findings we also highlight that we observe a switch-like behaviour in the case of short polymers. In this panel we report a typical event for a polymer $M=50$ beads long and with interaction strength $\epsilon=1.25 k_BT_L$ (see also Suppl. Movie M8). }
	\label{fig:switch}
\end{figure*}

\red{The results from the second study, (ii), are reported in Fig.~\ref{fig:firstorder}. In this figure we show a series of plots representing the joint probability distribution $P(R_g,\tilde{m})$, \emph{i.e.} the probability of observing the system in a certain state with given signed magnetisation $\tilde{m}$ and radius of gyration $R_g$. One may notice that the system undergoes a transition from a swollen (large $R_g$) and disordered ($\tilde{m}\sim 0$) phase to a compact (small $R_g$) and ordered (coherent magnetisation $\tilde{m} \simeq \pm 1$) one. In particular, at the transition point $\alpha_c=1.15$ (for $M=50$) the system shows the coexistence of both phases, \emph{i.e.} the probability has three maxima (as $T_L=T_{\rm Rec}$ this is an equilibrium model, hence, equivalently, the free energy has three minima). To gain these results, we have sampled the phase space near the critical point $\alpha_c$ as broadly as possible by performing $100$ independent simulations for a polymer of $M=50$ beads and runtime $10^6$ $\tau_{Br}$ each, from which we obtain the joint probabilities reported in Fig.~\ref{fig:firstorder}. Single trajectories of some of the $100$ runs are shown in Fig.~\ref{fig:switch} for the same values of $\alpha$ used for the joint probability plots.  }

\red{Finally, we highlight that we do not observe switching between the two symmetric metastable states, \emph{i.e.} $\tilde{m}=+1$ and $\tilde{m}=-1$, for a chain with $M=2000$ beads, but only for shorter chains (see Fig.~\ref{fig:switch} and Suppl. Movie M8). This switching property was reported in literature for effectively 1D models~\cite{Dodd2007,Sneppen2012,Jost2014}, where a relatively small number of nucleosomes were considered.}

\red{
This result is due to the fact that switching occurs when the system overcomes the energy barrier between the two states. This barrier grows with both the interaction strength $\epsilon$, and the number of intrachain interactions, which increases with $M$. In other words, the average first passage time from one state to the other can be predicted by a Kramers formula, so that it is proportional to the exponential of the free energy barrier, which scales with $M$, so that switching time increases exponentially with $M$ (or equivalently the switching probability decays exponentially with $M$). }

\vspace*{-0.4 cm}
\section{Contact Maps -- 2 State model}
\begin{figure*}[t]
	\includegraphics[width=1.01\textwidth]{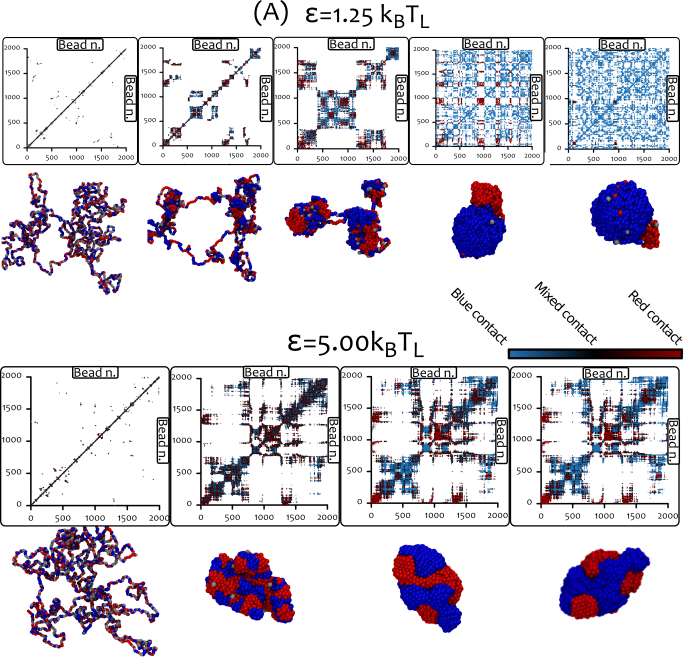}
	\caption{\textbf{Contact maps for the ``two-state'' model}. In this figure we show contact maps and representative snapshots corresponding to the dynamics of the system with two different values of $\alpha$. As one can notice, while low $\alpha=\epsilon/k_BT_L$ leads to a checker-board contact map at large times, high values of $\alpha$, or deep quenches, freeze the network of contacts. \red{Each point in the contact map is coloured red, blue or black if the entry in the matrix of contacts $C_{ij}$ is between two red beads, two blue beads, or a blue and a read beads, respectively. This is represented in the figure as a colour bar. Since the contact maps correspond to individual snapshots (i.e., they are not averaged over time), each bin is either coloured or empty. One can notice that high values of $\alpha$ (bottom row) leads to rapid folding of the chain and to the appearance of many mixed contacts (black points) which are then slowly lost (in favour of coloured contacts) over time.}}
	\label{fig:2SdbSI}
\end{figure*}

In Fig.~\ref{fig:2SdbSI} we report a series of contact maps for the ``two-state'' model, starting from the time at which the quench is performed. One can notice that, while for high values of the interaction parameter $\alpha$, the folding dynamics of the polymer, as well as the network of interactions, is frozen, for values of $\alpha$ closer to the transition point $\alpha_c=0.9$, the contact map evolves into a full checker-board interaction pattern. 

\vspace*{-0.4 cm}
\section{Decay of the Radius of Gyration}
In this section we illustrate a simple physical reasoning to rationalise the exponential decay of the gyration radius during the collapse at the transition point. Although there are some authors who argue that the collapse should be self-similar in time, and therefore, following a power law~\cite{DeGennes1985b,Abrams2002}, we have not found evidence of this self-similar collapse. 
This fact is presumably due either to the finite size of the chain used in our investigation, or to the initial condition. Indeed, in our simulations we start from random configurations far from a stretched coil, which is instead the situation often considered in theoretical models~\cite{DeGennes1985b}. Therefore in our case the common assumption of neglecting long-ranged loops at the early stages of the collapse~\cite{DeGennes1985b} may not be appropriate. Apart from the theory explored in Ref.~\cite{Kuznetsov1995}, we have not found in the literature a simple argument as to why the size of the polymer should decrease exponentially in time during the collapse. For this reason we illustrate a simple argument below. 

If one takes the growth (in number of monomers) of the pearls at very early times as $g \sim t^\beta$, with $\beta$ unknown for the moment, the volume of the pearls will grow as
\begin{equation}
	R^d_p \sim g^{d\nu} \sim t^{\beta \nu d}
\end{equation}
since each pearl is a crumpled globule $\nu=1/d$ and hence 
\begin{equation}
	R^d_p \sim g \sim t^{\beta}
\end{equation}
the total number of monomers in pearls is $g N_p$ (where $N_p$ is the number of pearls), therefore the number of inter-pearl monomers (not in the pearls) is
\begin{align}
	N_{ip} &= N - gN_p \sim N\left(1- g\dfrac{N_p}{N}\right) \\
	&\sim N\left( 1 - \dfrac{N_p}{N}t^\beta \right) \sim N e^{-N_p t^\beta/N}
\end{align}
as at early times $gN_p/N \ll 1$ and $t$ is small by definition of ``early-time''.
When pearls begin to appear, they are separated by a 3D distance given by the average number of inter-pearl monomers to the exponent $\nu$ and in particular the 3D distance $R_{ip}$ is
\begin{equation}
R_{ip} \sim \left(\dfrac{N_{ip}}{N_p}\right)^\nu \sim  \left(\dfrac{N}{N_p}\right)^\nu e^{-\nu N_pt^\beta/N}.
\label{eq:Rib}
\end{equation}
For $t=0$, Eq.~(\ref{eq:Rib}) correctly predicts that the typical size of inter-pearl distance is the whole polymer (as $N_p=1$). For $t\ne 0$, it predicts a stretched exponential decay of the gyration radius for $\beta<1$, and a simple exponential, for $\beta=1$. Therefore our argument provides a reason for a non-power-law decay of $R_g$.


We note that this argument is valid at very early times, or when the chain is large enough that the number of monomers belonging to the growing pearls  $N_p$ is much smaller than the number of monomers in the chain. It does not make any assumption regarding the presence of long range loops, while it makes the assumption that segments of the polymer not in pearls are still in a self-avoiding walk conformation ($R_{ib} \sim N_{ib}^\nu$). Although we have observed that the growing of pearls introduce competing tensions along the chain, at early times (or for very large chains), such forces do not spread across the whole chain, therefore leaving intra-blobs segments, tension-free.

Even if we cannot give an estimation for $\beta$ within our reasoning, this is not needed to prove the non-power-law decay of $R_g$ in time during the collapse. This exponent might assume values in between $\beta=1$ for a mean-field dynamics of a conserved order parameter~\cite{ChaikinLubensky} to $\beta \simeq 0.66$ as observed numerically for the coarsening of pearls during a homopolymer collapse~\cite{Byrne1995}. A more detailed study of the early stages of the collapse dynamics of a recolourable polymer might shed some light into the precise value of $\beta$ for this case, and on the precise nature of the decay of the radius of gyration (stretched versus simple exponential). 

\begin{figure*}[t]
	\centering
	\includegraphics[width=0.93\textwidth]{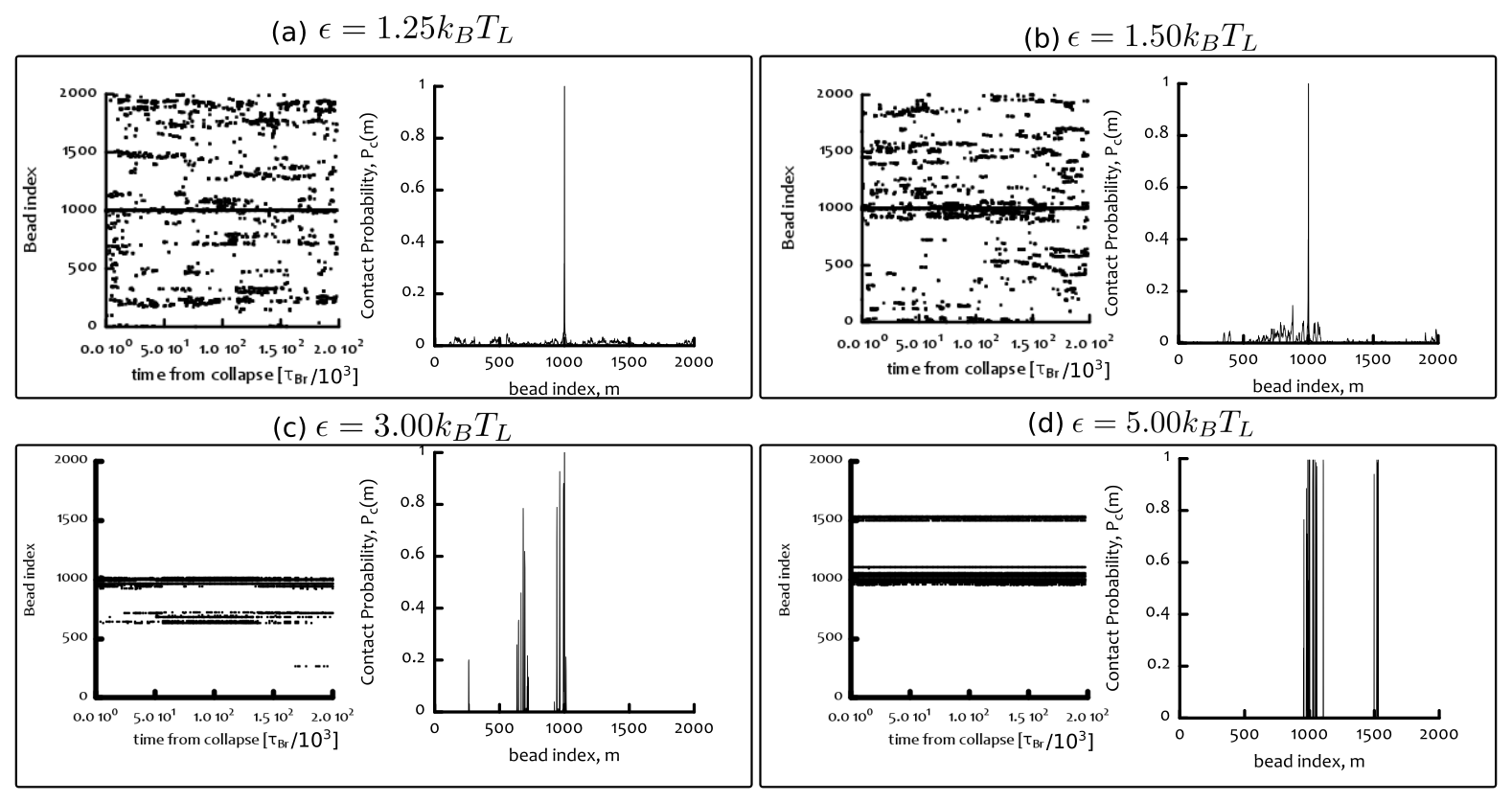}
	\caption{{\bf The folding of the polymer creates a quenched network of contacts.}  In this figure we show four pairs of graphs, each corresponding to a different choice of $\alpha$. 
For each pair, the left ``interaction'' plot is made by drawing a point for every bead $i$ which contacts bead $b=1000$ at a certain time-step after the collapse. As one can notice, these plots are highly dynamic (or ``fuzzy'') at low $\alpha$: this is because the network of contact is rearranging quickly during the simulation. Higher interaction strengths instead induce the selection of a subset of all possible interactions and create a frozen network of contacts. The right plots use the same information to show the time-averaged contact probability for the bead $b=1000$. For large enough $\alpha$, i.e. $\alpha>3$, these plots are clearly very different from the power-law decay of the contact probability $P_c(m)$ which is assumed in effectively 1D models (such a curve can only be recovered after averaging over many different runs, and different monomers, see Fig.~\ref{fig:Pc_ave}). }
	\vspace*{-0.4 cm}
	\label{fig:PC}
\end{figure*}

\begin{figure*}[h]
	\vspace*{-0.2 cm}
	\centering
	\includegraphics[width=0.7\textwidth]{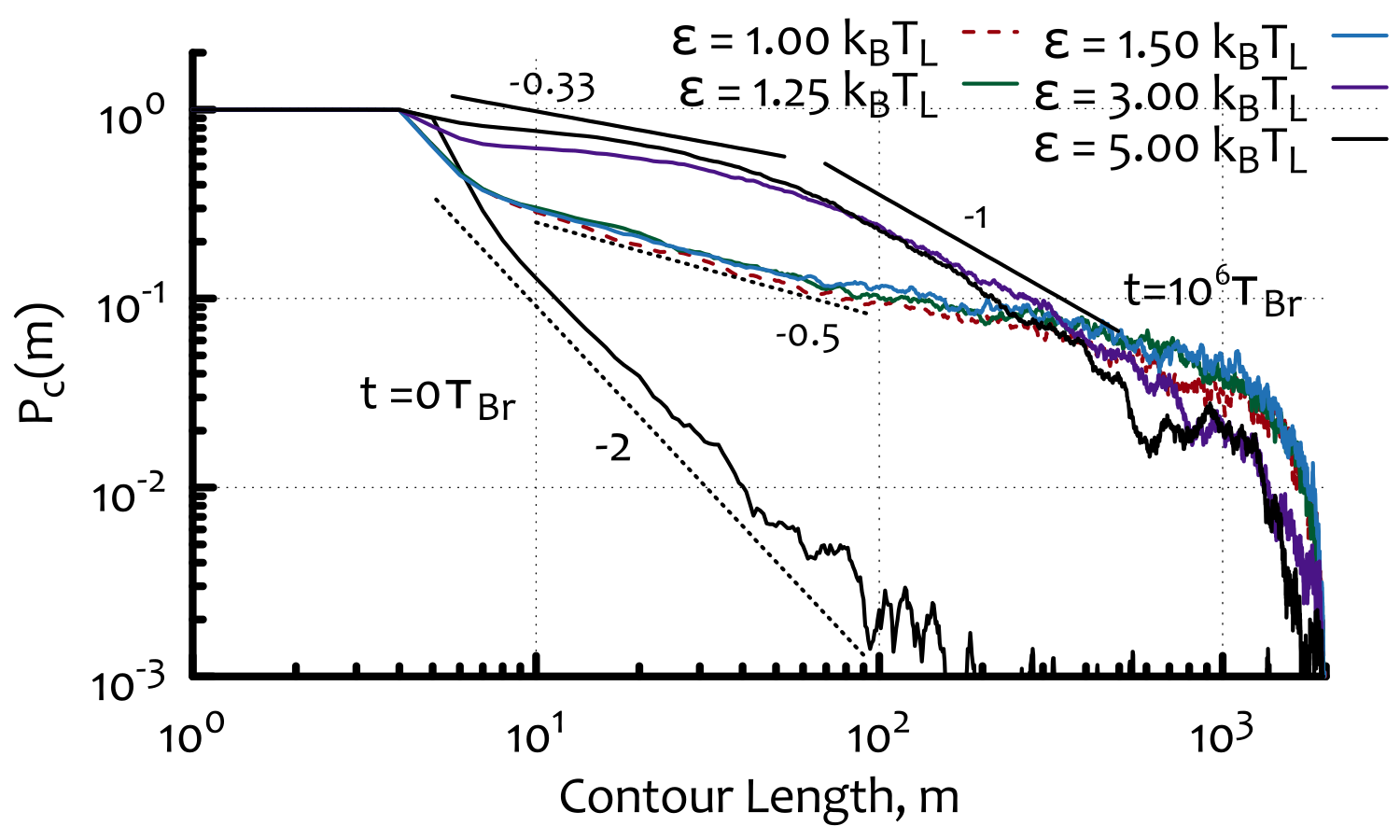}
	\caption{{\bf The contact probability averaged over independent replicas and beads leads to a power law statistics of contacts.} Here we show that a more standard contact probability curve emerges when it is averaged over different simulations, and over different monomers. The initial configurations show a very steep decay compatible with the self-avoiding walk statistics $c\simeq 2$ (for an ideal random walk one would have $c=1.5$ in 3d) while the collapsed states show $c\simeq 0.5$ for $1.0 \geq \epsilon/k_BT_L = \alpha \leq 1.5$. Higher interaction parameters lead to an enhancement of local contacts ($c=1/3$) followed by a steeper decay $c=1$ at longer ranges compatible with the fractal globule conjecture.}
	\label{fig:Pc_ave}
\end{figure*}

\begin{figure*}[t]
	\centering
	\includegraphics[width=0.9\textwidth]{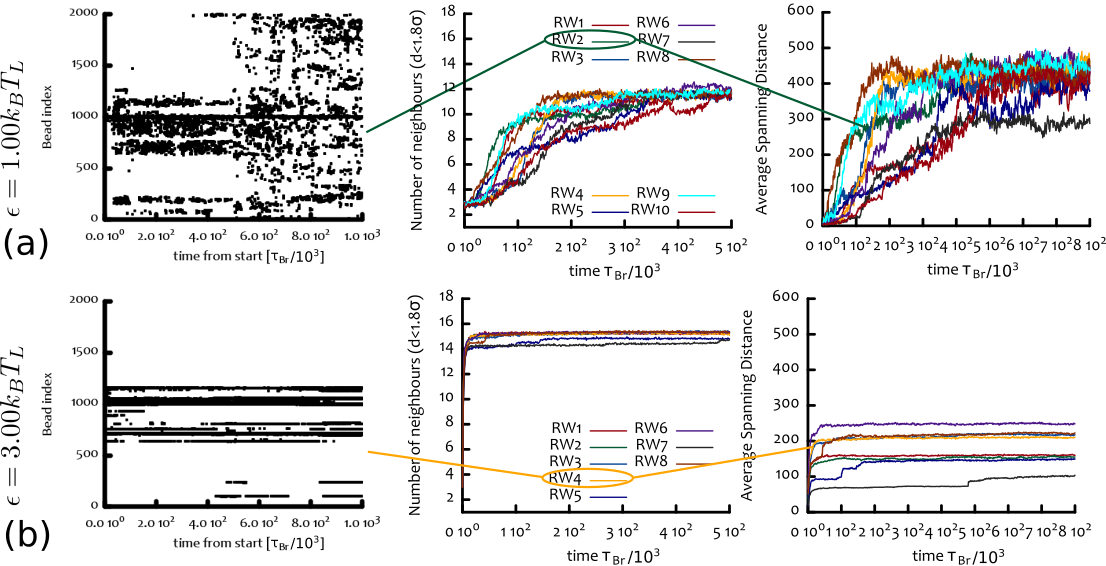}
	\caption{{\bf Dynamics of the contact networks: individual runs.} \textbf{(a,b)} From left to right, these panels show: the interaction plot (see Fig.~S6), the number of neighbours per bead over time, and the average spanning distance over time. The plots refer to three choices of $\alpha$ and selected simulations (denoted by ``RW'' and the index of the simulation). The choices of $\alpha$ correspond to \textbf{(a)} $\epsilon=k_BT_L$ and \textbf{(b)} $\epsilon=3 k_BT_L$. This figure shows that long range contacts develop dynamically during the collapse for $\alpha$ near the transition value $\alpha_c$, while for higher values of $\alpha$ the interactions are frozen (\emph{i.e.}, they do not evolve in time) and, although the number of neighbours is larger, the spanning distance is shorter, ultimately leading to slower epigenetic dynamics (see main text).}
	\label{fig:connectivity}
\end{figure*}

\begin{figure*}[t]
	\centering
	\includegraphics[width=0.85\textwidth]{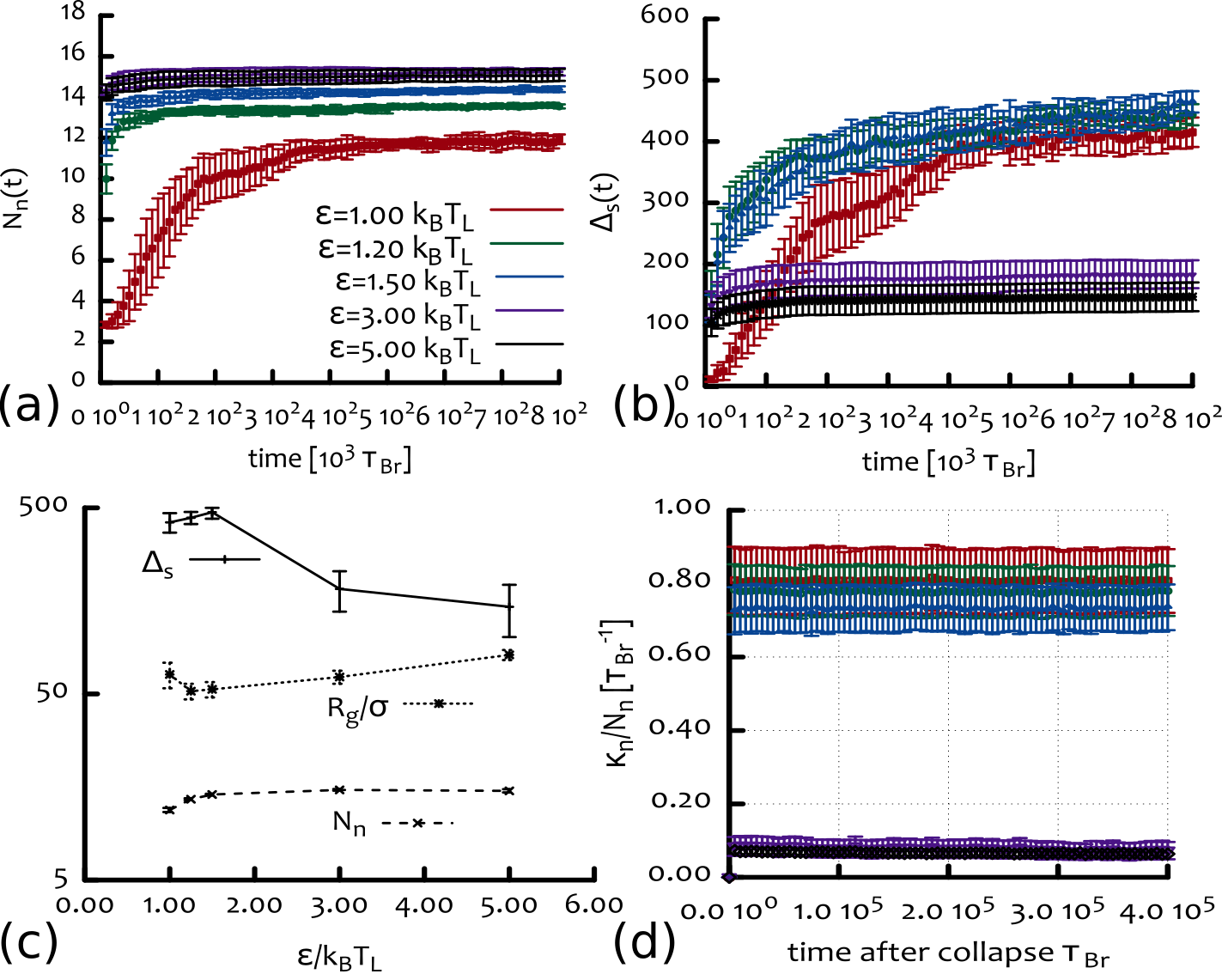}
	\vspace*{-0.5 cm}
	\caption{{\bf Dynamics of the contact networks: averages.} \textbf{(a)} Number of neighbours per bead, $N_n(t)$, averaged over different simulations. \textbf{(b)} Average spanning distance $\Delta_s(t)$ averaged over independent runs. \textbf{(c)} The values (averaged over last $5$ $10^5$ $\tau_{Br}$) of the number of neighbours, $\langle N_n \rangle$, and spanning distance $\langle \Delta_s\rangle$ plotted together with the average radius of gyration $R_g(t)/\sigma$ as a function of $\alpha$. \textbf{(d)} Plot of the fraction of exchanged neighbours per Brownian time $\kappa_n/N_n$ is here shown to achieve a steady state after the collapse. It can be seen that at large interaction strength the network of contacts displays a slower, more glassy dynamics than at low $\alpha$; furthermore at large $\alpha$ there are more short range contacts.}
	\label{fig:connectivity_ave}
	\vspace*{-0.5 cm}
\end{figure*}

\begin{figure}[t]
	\centering
	\includegraphics[width=0.4\textwidth]{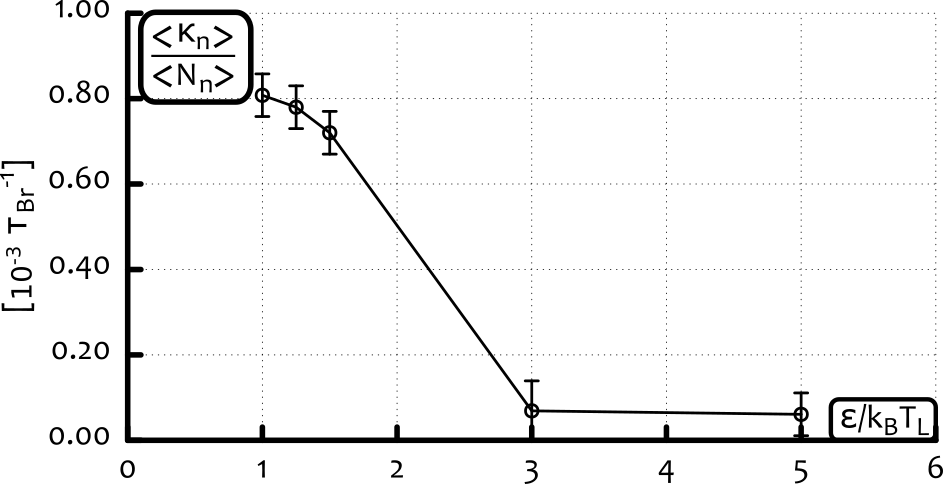}
	\vspace*{-0.2 cm}
	\caption{\textbf{The exchange of neighbours dramatically slows down at high $\alpha$'s.} In this figure we report the behaviour of $\kappa_n/N_n$ as a function of $\alpha=\epsilon/k_BT_L$. Thus, we show the fraction of neighbours exchanged per time-step on average by each bead. One can readily note a dramatic change in the  rate $\kappa_n$ when $\alpha \geq 3 k_BT_L$, for which case it attains a value of almost zero. This implies that the network of interactions is frozen, corresponding to a glassy exchange dynamics. }
	\label{fig:kappa_glass}
	\vspace*{-0.5 cm}
\end{figure}

\vspace{-0.3 cm}
\section{Contact Probability} 
\vspace{-0.3 cm}
In this Section we report the contact probabilities measured from our simulations  (see Figs.~\ref{fig:PC} and ~\ref{fig:Pc_ave}). 
In order to highlight the differences between our simulations and effectively 1D models we measured the contact probability of a single bead (index $b=1000$ along the chain, in Fig.~\ref{fig:PC}). While in effective 1D models one normally assumes a contact probability $P_c(m) \sim m^{-c}$ leading to long range interactions, which are tuned by $c$, here we observe that within within a single run the contact probability assumes a shape closer to a sum of delta-functions. 
This suggests a strong preferential selection of certain contacts along the polymer within that run, and a strong deviation from a ``mean-field'' view where contacts are regulated via $P(c)$ (which is an average over many conformations).
In Fig.~\ref{fig:PC} we report our findings by showing (in the left hand side graphs), an ``interaction plot'' which shows how the pattern of beads contacting bead $b=1000$ changes over time, for $\alpha=1.25 - 5$ $k_BT_L$. These plots show that the ``fuzziness'' that characterises mobile networks of contacts disappears when $\alpha \geq 3k_BT_L$. At these values of interaction strength, the contacts between some beads are present at all times and they never exchange. 

The plots in Fig.~\ref{fig:PC} also show (on the right hand side) the time averaged contact probability (again for bead $b=1000$). The graphs capture the strong departure from a mean-field-like interactions for high interaction strengths as in fact $P_c(m)$ resemble a sum of delta-functions rather than a power law. The picture that emerges is therefore similar to that of spin-like variables interacting on a network, where the edges are established by the collapse dynamics. When the interaction parameters are higher than a certain value the edges of the network are frozen in place, resembling a spin glass. We finally stress, that although we observe this departure from the mean-field assumption, the average of $P_c(m)$ over many beads and many simulations gives a more ``traditional'' power-law decay as we show in Fig.~\ref{fig:Pc_ave}. In particular, we find $P_c(m)\sim m^{-c}$ with $c$ ranging from $1/3$ to $1/2$ for different interaction strengths at the end of the collapse dynamics (see Fig.~\ref{fig:Pc_ave}), while they all start from a situation where $c \gtrsim 2$ compatible with a self-avoiding walk statistics (an ideal random walk would have $c=1.5$ in 3d). 

\vspace*{-0.6 cm}
\section{Connectivity}
\vspace*{-0.4 cm}
In this section we introduce and compute several quantities to characterise the change in network connectivity. As described in the main text, we track the average number of neighbours $N_n(t)$, and also the average spanning distance $\Delta_s(t)$ defined as
\begin{equation}
\Delta_s(t) = \dfrac{1}{M}\dfrac{\sum_{a \neq b} |a-b|P_{ab}(t)}{\sum_{a \neq b} P_{ab}(t)},
\end{equation}
where $a$ and $b$ denote polymer beads. The dynamical changes of these quantities during the collapse in selected runs (denoted with ``RW'' and the index of the simulation) are reported in Fig.~\ref{fig:connectivity}. From this figure it is important to notice that while there is an evident increase in number of neighbours and spanning distance for $\alpha=\epsilon/k_BT_L=1$, the same is not observed for higher interaction parameters. In these cases, \emph{e.g.}, the case with $\alpha=3$, the network of interactions is frozen, the number of neighbours quickly saturates to the maximum value while the spanning distance arrests to achieve a smaller values in steady state with respect to that for lower values of $\alpha$.  

The average of these quantities over different runs are shown in Fig.~\ref{fig:connectivity_ave}(a)-(b). Once again, we can readily see that while the number of neighbours increases and then plateaus at large $\alpha$, the spanning distance has a more complex dependence on the interaction strength. In Fig.~\ref{fig:connectivity_ave}(c) we also show the late time averaged values (taken after the collapse and over the last $5$ $10^5$ $\tau_{Br}$) alongside the value of the radius of gyration as a function of $\alpha$. One can notice that both $\Delta_s$ and $R_g$ are non-monotonic functions of $\alpha$, therefore suggesting the existence of a critical $\alpha_c$ above which the response of the system changes. In particular one can notice that for $\alpha \geq 3$ the spanning distance starts to decrease and the radius of gyration to increase, corresponding to the formation of more short ranged network and more frustrated configurations at higher $\alpha$. 

Finally in Fig.~\ref{fig:connectivity_ave}(d) we report the value of the neighbour exchange rate $\kappa_n$ divided by the average number of neighbours $N_n$ at any time after the collapse: it can be seen that $\kappa_n/N_n$ always reaches a steady state. This steady state value monotonically decreases and, in particular, undergoes a sharp transition around $\alpha_c \simeq 3$ (see Fig.~\ref{fig:kappa_glass}), above which the network of interactions is virtually frozen.

\green{
	\vspace*{-0.75 cm}
\section{Dependence on Initial Conditions within the ``two-state'' Model}
	\vspace*{-0.25 cm}
We note that the quenched disordered state observed for high values of $\alpha$  and identified as a ``topologically frozen'' state in the main text cannot be produced in the case the system was initialised as a uniformly coloured polymer (homopolymer). This initial condition would, in this case, lead to a standard homopolymer collapse. 
Another possible initial condition is a globular polymer with random colouring. This resembles the early stages of the collapse process shown in Fig.~S5 and Suppl. Movie M3 and we therefore expect that the subsequent evolution is very similar to the one investigated in the main text, where the polymer is initialised as swollen and disordered.
}

\newpage
\phantom{a}
\newpage
\phantom{a}
\newpage

\begin{figure*}[t]
	\centering
	\includegraphics[width=0.83\textwidth]{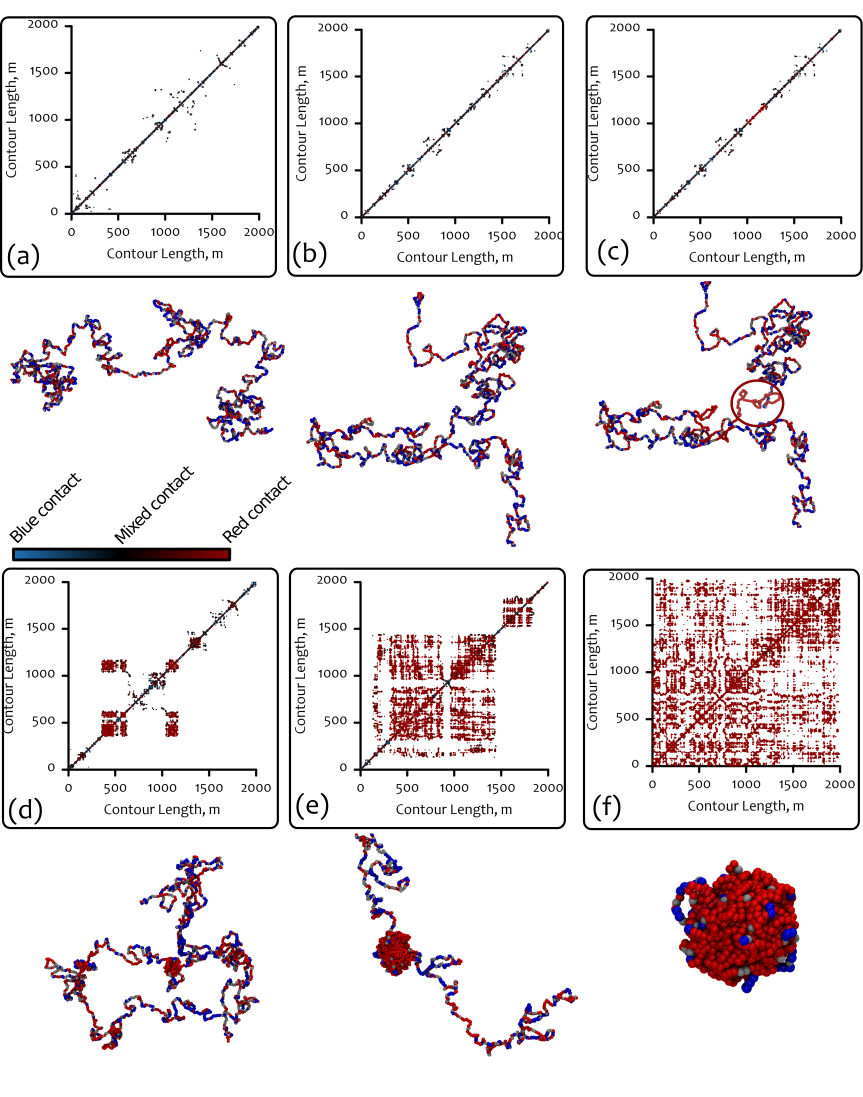}
	\caption{\textbf{Contact maps and snapshots for the ``two-state with intermediate state'' model before and after the perturbation.} In this figure we report the evolution of the contact map for the two-state with intermediate state model during the artificial recolouring. \textbf{(a)-(c)} refer to \textbf{(a)} $4$ $10^5$ $\tau_{Br}$ before the perturbation while \textbf{(b)} and \textbf{(c)} show the contact map immediately before and after the recolouring (in \textbf{(c)} (note the small red segment in the middle of the contact map). \textbf{(d)-(f)} refer to \textbf{(d)} $1$ $10^5$ $\tau_{Br}$, \textbf{(e)} $2$ $10^5$ $\tau_{Br}$ and \textbf{(f)} $4$ $10^5$ $\tau_{Br}$ after the perturbation. \red{The contact maps correspond to snapshots in time and each entry is coloured red, blue or black accordingly to whether the contact is red-red, blue-blue or mixed (see colour-bar in figure).} From this figure one can readily appreciate the dramatic change in conformation (phase transition) driven by the localised artificial recolouring. }
	\label{fig:2S+1dbSI}
\end{figure*}

\section{Contact Maps -- Two-State Model with Intermediate State}
In this section we report, in Fig.~\ref{fig:2S+1dbSI}, the contact maps for the ``two-state'' model with intermediate state (IS) at different times and during the local recolouring perturbation. We started from the mixed metastable state (MMS) and then forced the recoloured a small (10\%) segment in the middle of the chain. From Fig.~\ref{fig:2S+1dbSI} one can notice that this localised perturbation quickly drives the system towards the epigenetically coherent and globular state. As we show in the main text, this ordered phase is instead robust against major global reorganisation events such as a semi-conservative replication. This ultra-sensitive response, \emph{i.e.} a dramatically different response of the system to an external stimulus depending on the current state of the system, can be appreciated by looking at the large re-organisation and phase transition driven by such a small perturbation (Fig.~\ref{fig:2S+1dbSI} and main text Fig.~\ref{fig:2S+1db_MMS}) in contrast to the preserved coherent state even after a replication event (see main text Fig.~\ref{fig:2S+1db_MMS} at late time).

\section{Stability of Metastable Mixed State}
\begin{figure*}[t]
	\centering
	\includegraphics[width=0.9\textwidth]{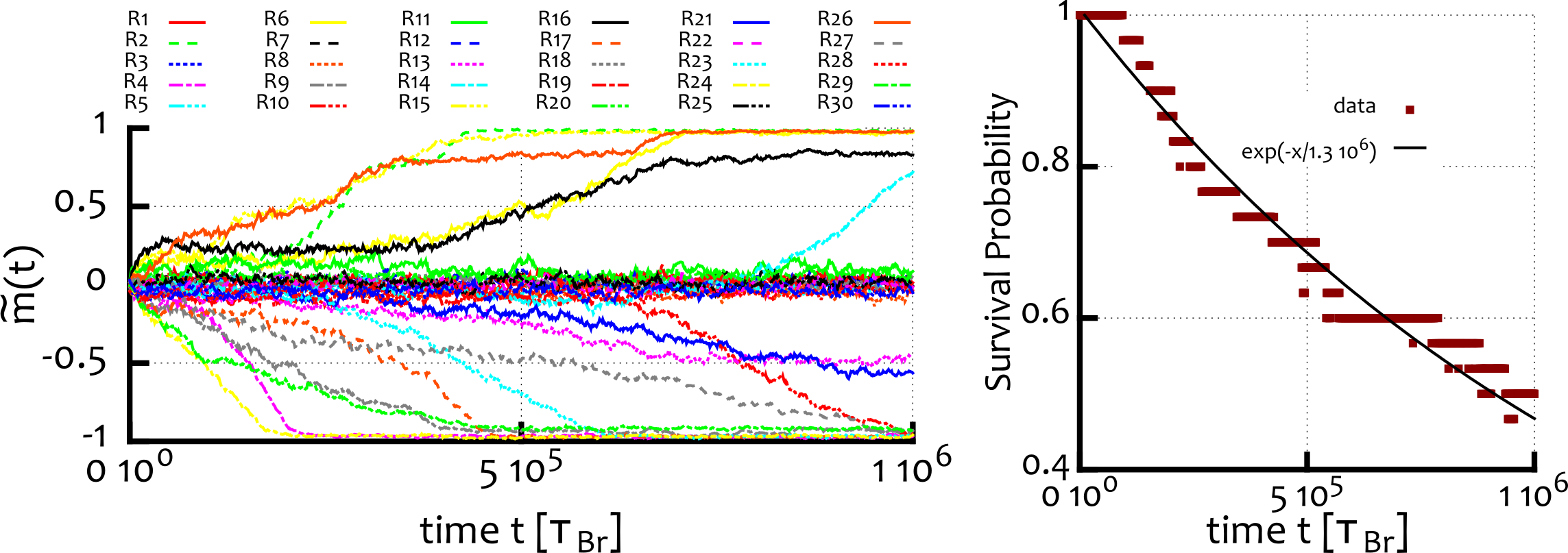}
	\caption{{\bf The survival probability of the MMS}. (left) Plot of the signed epigenetic magnetisation for the $30$ simulations performed. One can notice that a fraction of them ($50$\%) are still in the MMS state at the end of our simulation. (right) The survival probability can then be estimated by plotting the fraction of replicas still in the MMS at time $t$. This probability is here shown to exponentially decrease with a characteristic decay time of $1.3$ $10^6$ $\tau_{Br}$.}
	\label{fig:MMS_stab_SI}
\end{figure*}

\red{
In this section we aim to quantify the stability of the metastable mixed state (MMS) whose existence was reported in the main text in the ``two-state'' with intermediate state model. In order to do this we perform $30$ simulations with a chain $M=2000$ beads long and monitor the (signed) epigenetic magnetisation as a function of time. From this it is straightforward to obtain the survival probability of the MMS, i.e., the fraction of MMS which survives as a function of time. The results are reported in Fig.~\ref{fig:MMS_stab_SI} (for $\alpha=1$) where we show the behaviour of the signed epigenetic magnetisation $\tilde{m}$ for all $30$ simulations (left) as well as the survival probability (right). From the plot one can readily notice that at the end of the simulation (which last $10^6$ Brownian times, $\sim 3$ hours of real time, see mapping in Simulation Details) $50$\% of the replicas are still in the MMS. The survival probability can be fitted by a simple exponential with a characteristic  decay time $\sim 1.3$ $10^6$ Brownian times. It is also worth stressing that for the same value of $\alpha$, the simpler ``two-state'' model, shows instead that $100$\% of the simulations end up in the collapsed and epigenetically coherent state before $10^6$ Brownian times. This can be seen from Fig.~4a of the main text where we plot the average epigenetic magnetisation as a function of time: this is observed to saturate near unity at $10^6$ Brownian times for $\alpha=1$. We finally note that we observed that the MMS is unstable in the case $\alpha \geq 1.25$, as all the $10$ simulations we performed collapse within few hundreds of recolouring steps ($10^3$ Brownian times, not shown). On the other hand we have not thoroughly explored the parameter space in order to find the precise range of $\alpha$ for which the MMS is metastable.    
}

\newpage
\phantom{a}

\section{Theta point for a homopolymer}
In the main text we describe an argument for which our ``two-state'' model with broken detailed balance leads to long-lived TADs, \emph{i.e.} a block-like pattern along the contact map. As in this argument we use the value of the collapse transition temperature, or $\Theta$ temperature, for a homopolymer, here we estimate this approximately.

\begin{figure}[t]
	\includegraphics[width=0.5\textwidth]{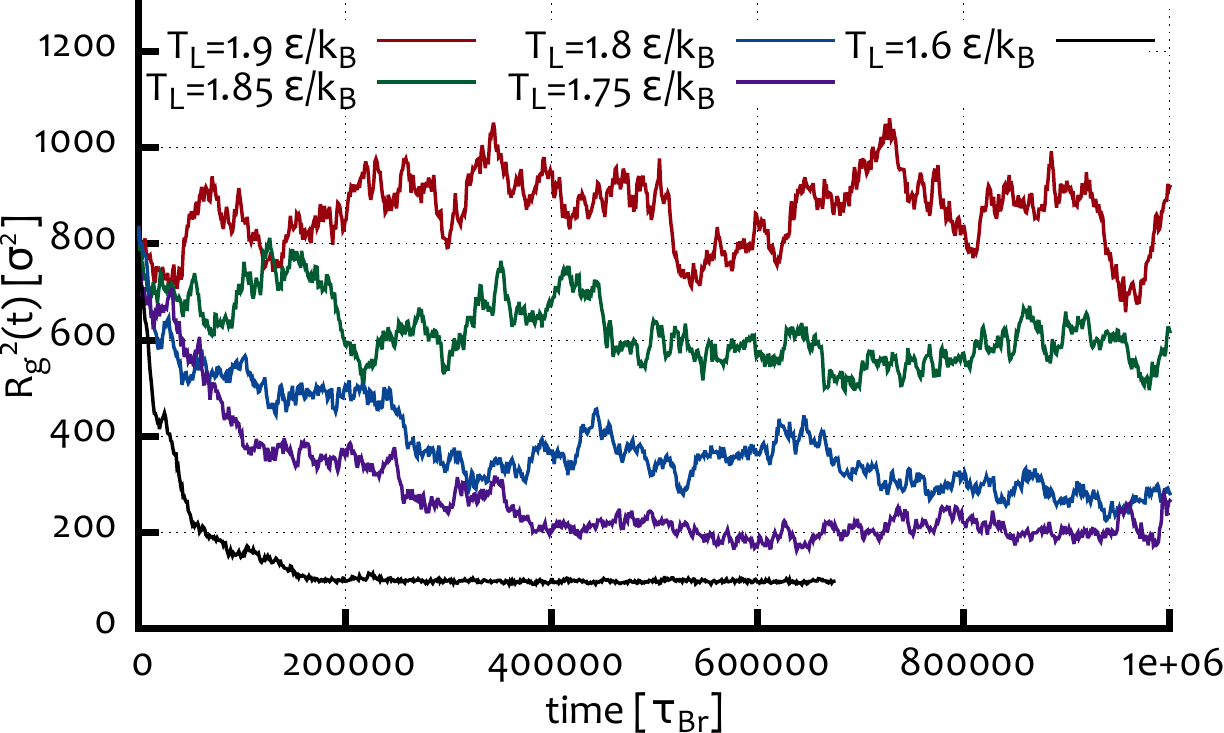}
	\caption{{\bf Theta-point for a non-recolouring homopolymer $M=2000$ beads long}. This figure shows the behaviour of the (squared) radius of gyration for a non-recolouring homopolymer starting from a self-avoiding swollen configuration. The curves are averaged over 10 independent simulations of the system. Temperatures above $T_L \simeq 1.85\epsilon/k_B$ do not appear to drive the collapse of the polymer and we therefore interpret this as an approximate estimate of the theta temperature of the system (at $M=2000$). }
	\label{fig:thetapoint}
\end{figure}

To do so, we study the behaviour of the radius of gyration for a homopolymer with $M=2000$ beads, for different temperatures $T_L$ and starting from a swollen self-avoiding configuration (Fig.~\ref{fig:thetapoint}). The curves are averaged over $10$ independent simulations. We observe that for $T_L \gtrsim 1.85\epsilon/k_B = T^{*,h}_L$ the polymer is not smaller than its initial state, and therefore consider this temperature as roughly the theta-point for the homopolymer (non-recolouring) case.
At this stage it is worth reminding that the critical point for an equally long ``recolourable'' polymer is at $T^{*,r}_L \simeq 1.11 \epsilon/k_B$.  

\green{
\vspace*{-0.5 cm}
\section{Dependence on Initial Conditions and Semi-Conservative Replication Protocol within the ``two-state with intermediate state'' Model}
The mixed metastable state (MMS) that is observed for values of $\alpha$ close to the critical value $\alpha_c$ is not observed for higher values of $\alpha$. This is most likely because this MMS becomes unstable. Moreover, we have never observed the MMS emerging from polymers which were initially homogeneously coloured (homopolymers). On the other hand, we have observed the onset of MMS from a disordered and collapsed polymer. 
Within the same ``two-state with intermediate state'' model, in the main text we also report a case in which the polymer is stable against semi-conservative replications which turn 50\% of the beads into a random colour. We also studied (data not shown) the case in which 50\% of the monomers are turned grey (inactive). The collapsed state is stable also against this type of perturbation. It is a future challenge to investigate biologically-relevant replication strategies that might be overturn the dominant epigenetic mark and force an epigenetic switch in the system.    
}

\section{Contact Maps -- Two-State Model with Broken Detailed Balance}
In this section we report selected instantaneous contact maps for the ``two-state'' model with broken detailed balance. 
As one can notice, the off-diagonal contacts are temporary and very mobile. Averaging over such frames leads to the ``block-like'' structure reported in the main text, which is reminiscent of the TAD-like structures often reported by capture experiments in eukaryotic cells.   

\begin{figure*}[t]
	\includegraphics[width=\textwidth]{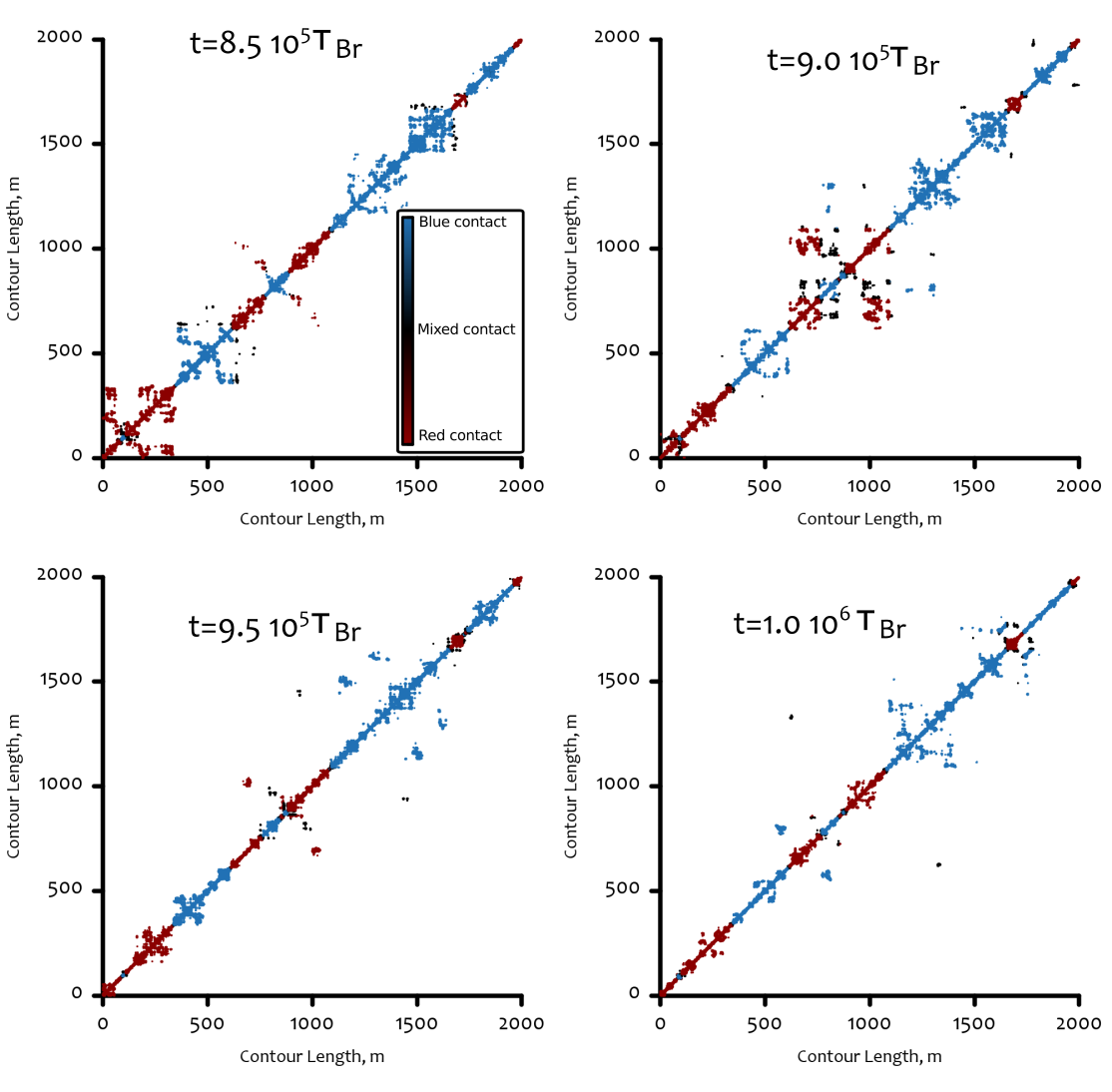}
	\caption{\textbf{Snapshots of the contact maps for the ``two-state model'' with broken detailed balance model}. This figure shows the instantaneous contact maps at different timesteps within the last 200 recolouring steps. \red{The colour code for the maps is based on the colours of the beads forming the contact: red, blue or black for red-red, blue-blue or mixed contacts, respectively (see colour bar)}. From this figure one can appreciate the temporary aggregation of globules, whose underlying 1D epigenetic segment is forming long-lived domains during the simulation. Temperatures were: $T_L= 1.75 \epsilon/k_B$, $T_{\rm Rec}=0.1 \epsilon/k_B$.}
	\label{fig:2SNOdb_cmapsSI}
	\end{figure*}
	
	\section{The Epigenetic Correlation Length}
	In analogy to the Ising 1D magnetic correlation length, the ``epigenetic correlation length'', can be measured by imagining that each bead has a epigenetic state ${q}$ that can take any of the possible states: red, blue or grey. The epigenetic correlation function can therefore be expressed in terms of these variables as 
	\begin{equation}
	g_{1D}(m) \equiv \langle \delta_{{q}(0),{q}(m)}\rangle 
	= \left\langle \dfrac{1}{M-m} \sum_{ij} \delta_{q_i q_j}\delta(|i-j|-m) \right\rangle,
	\end{equation}
	and where the average is performed over independent system configurations and over different (uncorrelated) times. Since it is analogous to the 1D Ising (or Potts) magnetisation correlation function, $g_{1D}(m)$ takes the functional form
	\begin{equation}
	g_{1D}(m) \sim e^{-m/\xi(T_{\rm Rec})}
	\end{equation}
	where the correlation length $\xi(T_{\rm Rec})$ diverges with $M$ as the system breaks the symmetry between red/blue epigenetic states. When the system displays different coexisting epigenetic states (multi-state regime), either in the  glassy phase of the simplest ``two-state'' model or in the stable ``block-like'' organised regime of the model with broken detailed balance, $\xi(T_{\rm Rec})$ is finite. 
	
	This length-scale is clearly dependent on $T_{\rm Rec}$: even if the polymer were an immobile straight line, the correlation length would increase with decreasing $T_{\rm Rec}$ and ultimately diverge as $T_{\rm Rec}\rightarrow 0$ (as there is no transition in 1D Ising-like models with short-range interactions). Furthermore, in our model the dynamics of the polymer is coupled to the epigenetic organisation, and therefore $\xi(T_{\rm Rec})$ is expected to depend more subtly on $T_{\rm Rec}$ (as well as $T_L$). 
	
\vspace*{-0.3 cm}
\section{Other regimes of the ``two-state'' model with broken detailed balance}
\vspace*{-0.3 cm}
\red{	
In this Section we report other regimes that we observe in the ``two-state'' model with broken detailed balance. In particular, we fix $T_{\rm Rec}=0.1 \epsilon/k_B$ and change the value of $T_L$ in the range $T_L \in [1.5, 6] \epsilon/k_B$.  We observe that $T_L \leq 1.75 \epsilon/k_B$ leads to a collapse of the coil into a single-state dominated globule, similar to that observed in the standard ``two-state'' model, but with a higher critical temperature (recall this is $1.11 \epsilon/k_B$ for the case with $T_L=T_{\rm Rec}$). 
For the temperature range we considered, we found a transient, long-lived regime, where the averaged contact map a characteristic ``block-like''. Here, higher temperatures $T_L$ lead to a smaller size of the domains for fixed and small $T_{\rm Rec}$ (the domains are barely visible for $T_L=6$ $\epsilon/k_B$).  
The contact maps reported in Fig.~\ref{fig:2SNOdb_mix} show the average frequency of contacts (upper triangle), and a weighted contact map (lower triangle), where each entry is coloured based on the frequency of contacts through the following method
$$C_{ij} =\dfrac{\sum^T_{t=0}\Theta(-r_{ij}(t)+R_i)F(q_i,q_j)}{\sum^T_{t=0}\Theta(-r_{ij}(t)+R_i)}. $$
In this equation, $F(q_i,q_j)=1$ if $q_i=q_j=1$ (blue), $F(q_i,q_j)=-1$ if $q_i=q_j=2$ (red) and $F(q_i,q_j)=0$ if $q_i\neq q_j$; while $\Theta(x)=1$ if $x>0$ and $0$ otherwise and $R_i=1.8\sigma$ (see main text).
By weighting the contacts in this way, one can classify the entries of the contact matrix based on the frequency of the interactions between same-coloured beads; this helps distinguish the different epigenetic domains along the polymer. In other words, entries $C_{ij}$ in the contact map that are fully red (blue) indicate that all observed contacts between $i$ and $j$ were between red (blue) beads. The observed pattern is not trivial since the underlying epigenetic landscape is dynamic, therefore in principle a bead can become blue even though belonging to a red domain. Note that by using the same normalisation as for a standard contact map (maximum number of observed contacts) would lead to less well defined (whiter) domains. Thus, in Fig.~\ref{fig:2SNOdb_mix} one can readily appreciate the existence of epigenetically stable domains that show enhanced contacts within themselves and little mixed interactions with nearest neighbouring ones.
\begin{figure*}[t]
\centering
\includegraphics[width=0.7\textwidth]{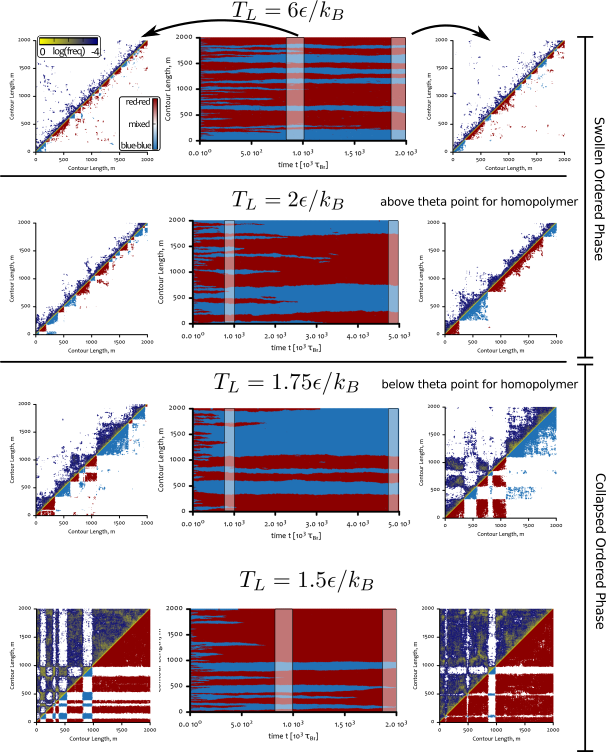}
\caption{\red{{\bf Classification of the regimes in the ``two-state'' model with broken detailed balance}. This figure shows the observed regimes for the two-state model with broken detailed balance. Simulations are initialised with the system in the swollen disordered phase. Here, we fix $T_{\rm Rec}=0.1 \epsilon/k_B$ and vary (from bottom to top) $T_L \in \{ 1.5, 1.75, 2, 6\} \epsilon/k_B$. We observe that $T_L \leq 1.75 \epsilon/k_B$ leads to a collapse of the coil into a single-state dominated globule, similar to that observed in the standard ``two-state'' model, but with a higher critical temperature. On the other hand, setting $T_L > 1.75 \epsilon/k_B$ eventually leads to a swollen ordered regime in steady state (see Fig.~\ref{fig:NOdb_ic}). All simulations show an interesting long-lived transient regime, where local domains coexist along the chain giving a characteristic ``block-like'' pattern to the averaged contact map.}}
\label{fig:2SNOdb_mix}
\end{figure*}
}
	
	\newpage
\section{Stability of the epigenetic domains in the out-of-equilibrium limit of the model}
\begin{figure*}[t]
\centering
\includegraphics[width=1.0\textwidth]{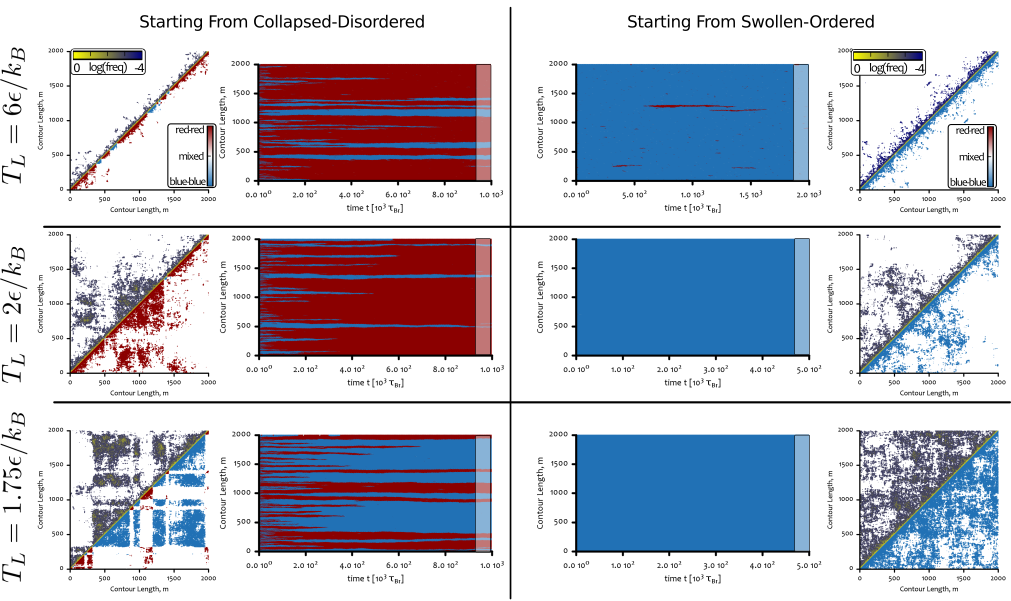}
\caption{{\bf Stability of the epigenetic domains ``two-state'' model with broken detailed balance: effect of initial conditions}. (Left) Collapsed disordered initial condition. Here epigenetic domains form and coalesce slowly over time. The contact maps show that the chain opens up significantly for the higher temperatures. (Right) Swollen ordered initial conditions. Here the single domain is stable, and the polymer undergoes a standard homopolymer collapse transition.}
\label{fig:NOdb_ic}
\end{figure*}

\red{Here we comment on the stability of the epigenetic domains that can be seen emerging in the out-of-equilibrium limit of the ``two-state'' model when the (Langevin) temperature is larger than the theta temperature at which the polymer collapses when $T_P=0.1 \epsilon/k_B$ (about $1.75 \epsilon/k_B$). Here we address the question of what happens when the initial configuration with different initial conditions, either collapsed disordered (left in Fig.~\ref{fig:NOdb_ic}) or swollen ordered (right in Fig.~\ref{fig:NOdb_ic}).
For the first case, as in the main manuscript, TAD-like long-lived domains form, and tend to coalesce on a very long time-scale (corresponding to hours of physical time).}
\red{Starting from the swollen ordered phase, the single epigenetic domain is stable and no other ones form over the course of the simulations. As for the case of the homopolymer, the chain remains swollen for $T_L=2 \epsilon/k_B$, and eventually collapses for $T_L=1.75 \epsilon/k_B$.}

\end{document}